\documentclass[11pt]{article}

\usepackage[margin=1in]{geometry}

\usepackage{amsmath, amsfonts, amssymb, amsthm}
\usepackage[authoryear,round]{natbib}
\usepackage{graphicx}
\usepackage{microtype,xurl}
\usepackage{bm,tikz}
\usepackage{float,placeins}
\usepackage{thm-restate}
\usepackage{thmtools}
\usepackage[
  colorlinks=true,
  linkcolor=black,
  citecolor=black,
  urlcolor=blue!65!black,
  bookmarks=false,
  hypertexnames=false
]{hyperref}

\usepackage{multirow}

\usepackage{cleveref}
\crefname{equation}{Eq.}{Eqs.}
\Crefname{equation}{Eq.}{Eqs.}
\crefname{subequation}{Eq.}{Eqs.}
\Crefname{subequation}{Eq.}{Eqs.}
\crefname{figure}{Fig.}{Figs.}
\Crefname{figure}{Fig.}{Figs.}
\crefname{appendix}{App.}{Apps.}
\Crefname{appendix}{App.}{Apps.}
\usepackage{algpseudocode}
\usepackage{indentfirst}

\usepackage{makecell}

\renewcommand\cite[1]{\citep{#1}}

\usepackage{xcolor}

\usepackage{censor}

\usepackage{booktabs} \par

\usepackage{color}
\usepackage{caption}
\usepackage{subcaption}

\usepackage[group-separator={,}]{siunitx}
\newcommand\blfootnote[1]{	\begingroup
	\renewcommand\thefootnote{}\footnote{#1}	\addtocounter{footnote}{-1}	\endgroup
}

\usepackage{algorithm2e}

\SetKwInput{KwParams}{Hyperparameters}
\SetKwInput{KwVars}{Variables}

\bibpunct[, ]{(}{)}{,}{a}{}{,}
\AddToHook{file/appendix_risk_priority.tex/before}{\begingroup\let\small\footnotesize\setlength{\tabcolsep}{3pt}}
\AddToHook{file/appendix_risk_priority.tex/after}{\endgroup}

\title{Redesigning Service Level Agreements:\\Equity and Efficiency in City Government Operations}
\author{
Zhi Liu\\
Cornell Tech\\
\texttt{zl724@cornell.edu}
\and
Nikhil Garg\\
Cornell Tech\\
\texttt{ngarg@cornell.edu}
}
\date{}

\begin{document}

\maketitle

\begin{abstract}
We study how government should allocate resources (e.g., maintenance of public infrastructure) over time, efficiently and equitably -- when these desiderata may conflict. In particular, we consider the design of Service Level Agreements (SLAs) in city government operations: promises that incidents such as potholes and fallen trees will be responded to within a certain time.

We model the problem of designing a set of SLAs as an optimization problem with equity and efficiency objectives under a queueing framework; the city has two decision levers: how to allocate response budgets to different neighborhoods, and how to schedule responses to individual incidents. We first theoretically analyze a stylized model, deriving when the equity-efficiency trade-off is substantial in absolute terms and centralizing response efforts leads to relatively smaller improvements over optimal decentralized operations. We then develop a simulation-optimization framework to design and analyze policies in practice. Applying our framework to data from NYC, we find: (a) in practice, despite large potential trade-offs, equity improvements are possible with relatively small decreases in efficiency; (b) there exist policies that are both more efficient and equitable than status quo inspections; (c) the gain from centralization is {relatively small compared to the potential efficiency-equity tradeoff}.

Our results provide widely applicable predictions about the scale of the equity-efficiency trade-off in similar resource allocation settings, and our data-driven simulation-optimization framework enables specific policy recommendations in government operations. Our code is available at \url{https://github.com/nikhgarg/sla_design}. Our theoretical statements have been formalized in Lean, available at \url{https://github.com/nikhgarg/AppliedModelingLib/blob/main/papers/LG24ServiceLevelAgreements/}. 

\end{abstract}
\noindent\textbf{Keywords:} Service Level Agreement, government operations, equity-efficiency

\blfootnote{
			We thank the New York City Department of Parks and Recreation for their valuable work and inside knowledge, and we especially thank Uma Bhandaram and Tyler Gibson. We also benefited from discussions with Anum Ahmad. This work was supported in part by the Cornell Tech Urban Tech Hub, Amazon and Meta research awards, and an NSF CAREER Grant \#2339427. This paper previously appeared in the Twenty-Fifth ACM Conference on Economics and Computation (EC'24) as an extended abstract.
	}

\section{Introduction}
\label{sec:intro}
\noindent Municipal government agencies, along with other organizations, must make time-sensitive allocation decisions (such as fixing potholes, collecting garbage, restoring power, and making emergency repairs) and communicate their performance to stakeholders (such as the public and elected officials). Performance guarantees often take the form of \textbf{Service Level Agreements} (SLAs): incidents of type $k$ in neighborhood $b$ will, with probability at least $\beta$, be addressed within $z_{k,b}$ days: 
\[
    \Pr\!\left(T_{k,b}\leq z_{k,b}\right)\geq \beta.
\]
For example, New York City -- our data setting -- has long required agencies to establish such SLAs for 311 service-request categories; NYC publishes these commitments, and compliance (the percentage of requests meeting planned times) is supposed to be tracked annually.\footnote{New York City has required agency-specific 311 SLAs \href{https://www.nyc.gov/assets/records/pdf/executive_orders/2008EO115.pdf}{since 2008}; current law requires NYC311 to \href{https://codelibrary.amlegal.com/codes/newyorkcity/latest/NYCadmin/0-0-0-114437}{publish agency SLAs online}; and request-level records include the corresponding \href{https://data.cityofnewyork.us/api/views/erm2-nwe9}{due dates}. The FY2025 \href{https://www.nyc.gov/assets/operations/downloads/pdf/mmr2025/dpr.pdf}{Mayor's Management Report}, for example, reports NYC Parks' annual percentage of damaged-tree requests meeting an eight-day time to first action (p.~164) (links in-line).}
According to these public SLAs, for example, our practitioner collaborators, the \textit{Department of Parks and Recreation (NYC Parks)}, will respond to a report of \textit{Illegal Tree Damage} within \textit{8 days}. Other cities -- such as Houston, Toronto, Boston, Brisbane, San Francisco, and Washington, DC -- all provide similar service guarantees for various incidents, occasionally with publicly available dashboards or reports tracking compliance; e.g., San Francisco and Philadelphia aim to repair potholes within 3 days 90\% of the time, and Brisbane aims to investigate ``urgent'' issues within 24 hours and non-urgent ones within a week.\footnote{See \href{https://www.houstontx.gov/policies/adminpolicies/2-23.pdf}{Houston}, \href{https://www.toronto.ca/wp-content/uploads/2025/10/8fda-311-Service-Standard-Dashboard-2025-Q1-Q2.pdf}{Toronto}, \href{https://www.boston.gov/departments/analytics-team/cityscore}{Boston}, \href{https://www.brisbane.qld.gov.au/online-services/report-an-issue/potholes}{Brisbane}, \href{https://sfpublicworks.org/calendar/san-franciscos-road-conditions-improve-ranked-highest-large-bay-area-cities-annual-roads}{San Francisco}, \href{https://ddot.dc.gov/service/pothole-repair}{Washington, DC}, and \href{https://philly-stat-360.phila.gov/pages/potholes}{Philadelphia}. Toronto also publishes \href{https://www.toronto.ca/legdocs/mmis/2025/bu/bgrd/backgroundfile-252536.pdf}{annual service-level tables (p.~76)} (links in-line).}
Similar guarantees are used to characterize and communicate time-sensitive performance in cloud computing \citep{patel2009service}, various web services \citep{jin2002analysis}, and ticketing systems broadly. This ubiquity reflects SLAs' desirable properties: (a) they are \textit{transparent} and externally auditable; and (b) if designed properly, they can translate to equity and efficiency goals: more urgent incidents should have shorter response timelines, and overall (importance-weighted) delays should not be disparate across areas. 

\begin{figure}[tb]
    \centering
    \includegraphics[width = .35\textwidth]{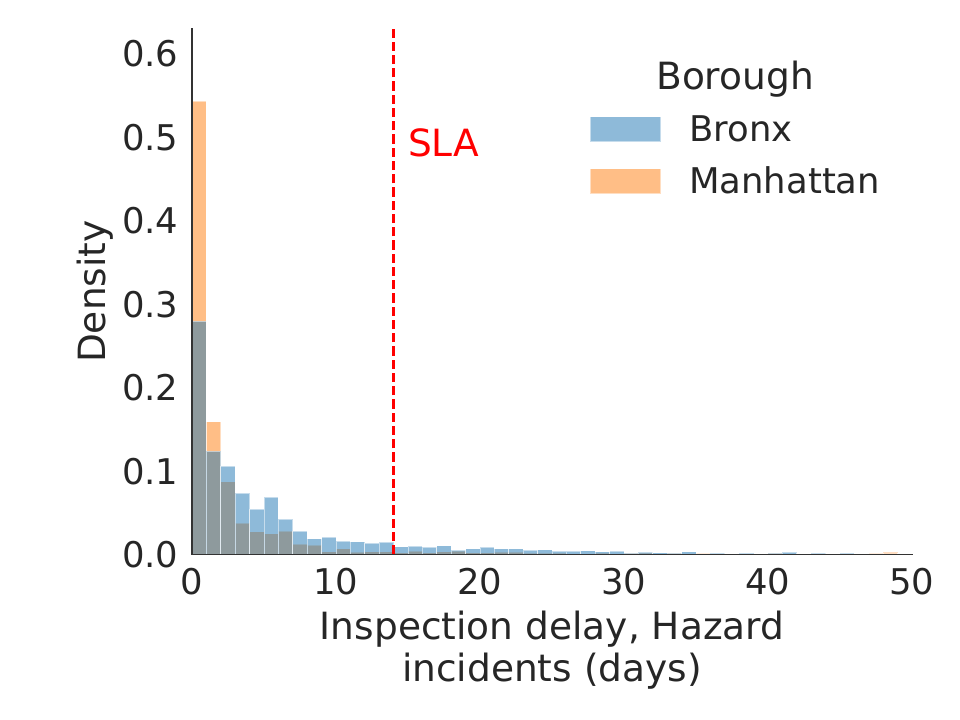}
    \includegraphics[width = .35\textwidth]{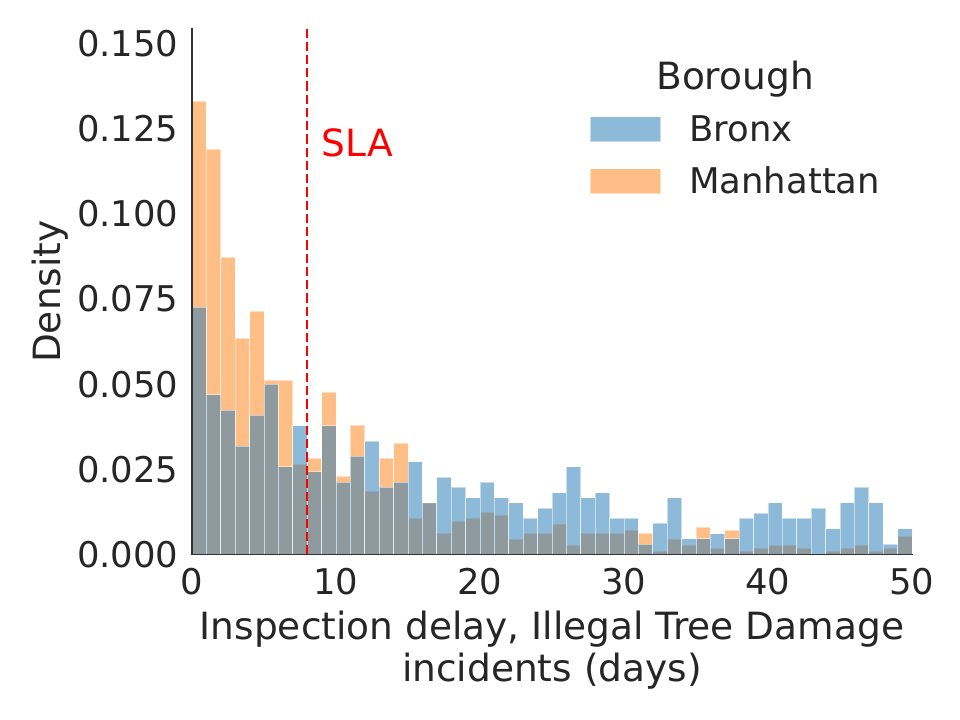}
    \caption[]{The Department of Parks and Recreation of NYC responds to service requests on ``Hazard'' and ``Illegal Tree Damage'' events related to street trees, among others. We find that even conditional on the same category of incidents, the distribution of the inspection delay (in this case defined by the time from the first service request to the completion of an inspection) varies by Borough. Moreover, a substantial number of service requests are not responded to within the publicly available SLAs, with some others not inspected at all. The data-driven distribution of the inspection delays also does not correspond to the priority defined in the SLAs: though hazard incidents have a looser SLA compared to illegal tree damage incidents (14 days versus 8 days), they are generally inspected sooner, reflecting their higher average risk rating. These inspection delays are calculated for requests created from 2019 through 2022 using public data, available at \url{https://data.cityofnewyork.us/Environment/Forestry-Inspections/4pt5-3vv4}.}
    \label{fig:hazardhist}
\end{figure}

How should SLAs be set? On the one hand, SLAs must correspond to capabilities: what resources does an agency have, and are they sufficient to meet a given set of SLAs? In other words, agencies meet SLAs as a function of their allocation decisions, not directly. Thus, one common approach is to design SLAs based on historical performance; by definition, such SLAs were achievable using historically available resources.\footnote{\href{https://www.houstontx.gov/policies/adminpolicies/2-23.pdf}{Houston's citywide service-request policy}, for example, directs the Citywide 311 Performance Manager to set each request-type SLA at a level departments can achieve 90\% of the time, using historical periods that represent demand and available resources and preferably at least one full year of data (Section~5.1.1, p.~2).}
On the other hand, SLAs are not neutral guarantees: they are high-dimensional, joint decisions that encode policy decisions and tradeoffs. Given a fixed overall budget, more effective service for one type or one neighborhood must trade off with service of another kind. Merely encoding historical allocation decisions thus ignores the managerial question of the desirability of these decisions. Thus, an alternate approach is to set them using political processes, weighing competing priorities; however, this approach may lead to SLAs that are unachievable using available resources. Communication with NYC Parks indicates that current SLAs are too inaccurate to meet or guide operations. \Cref{fig:hazardhist} shows publicly listed SLAs for two types of incidents, alongside how quickly these incidents were responded to in two Boroughs.\footnote{We consider the allocation of \textit{inspections} in response to requests, and so interchangeably use ``response'' and ``inspection.''}

\paragraph{Conceptual contribution.} At a high level, we turn SLA design from a descriptive question -- what SLAs do prior allocations support? -- to a prescriptive one -- what should SLAs and allocations be, to support given policy goals, while being achievable? Effective design must both reflect operational constraints and policy objectives. To do so, we transform SLA design into a task of optimizing two levers: (1) {\text{budgets}} in each neighborhood (i.e., number of workers who can respond to incidents) and (2) {allocation guidelines} for how workers prioritize incidents of different types.\footnote{Why these levers, instead of directly optimizing incident-level decisions? Spatial budget levels describe the status quo, for both administrative and logistical reasons: worker home offices are distributed throughout the city as determined by the budgets, and it is more efficient for a worker to respond to spatially nearby incidents. Thus, these levers \textit{complement} daily incident-level decision optimization (which specific open incidents should a worker address that day), determining the feasibility of a specific daily decision (the agency cannot easily inspect more incidents in a neighborhood than its worker capacity and spatial distribution allow).} As we show, these levers alongside information on incident arrivals directly translate to SLAs in a natural queueing model. Using this model, we both conceptually analyze the design space and engineer SLAs in a data-driven manner. Motivated by conversations with practitioners, we address two managerial and policy questions. (1) Is there a large equity-efficiency tradeoff, in which addressing incidents quickly overall may lead to inter-neighborhood disparities? Can current practice be improved on both efficiency and equity metrics? (2) What is the benefit, in terms of both efficiency and equity, of centralizing resources so that they can be dispatched across neighborhoods? Can we improve SLAs uniformly by doing so? How do the answers to both questions depend on the problem setting?

\paragraph{Technical approach.}

We take a three-step approach to answer the questions theoretically and through data. We (a) formulate and analyze a stylized queueing model, which induces a tractable optimization problem to determine budgets and incident type prioritization, giving insights on the price of equity and gain from centralization; (b) extend the model to incorporate real-world complexities, which can be optimized using a simulation-optimization framework; (c) optimize policies in an extended case study, using actual incident and inspection data from NYC Parks.

\paragraph{(a) A theoretical model to analyze SLAs.} We model SLA design as an optimization problem under a queueing network framework, with decision levers: allocating inspection budgets (inspectors) to different neighborhoods and scheduling priorities for different types of incidents. We formulate a convex optimization problem and simultaneously solve for an optimal budget allocation plan and inspection scheduling prioritization policy, to find the best feasible SLA under some objective function. We show that our model is general enough to allow a large class of objective functions, encoding both equity and efficiency. Our model further accommodates other administrative policies, such as when the city maintains a centralized inspection team. We theoretically study this model under a specific class of objectives, where the efficiency loss is a risk-weighted incident cost across the city (including both a noninspection cost and a cost for the tail delay implied by the SLA), and the equity loss measures {within-category differences in costs across neighborhoods}. Conceptually, we characterize the trade-off between equity and efficiency and the conditions under which it is substantial {in absolute terms} -- when different Boroughs face sufficiently different situations, and when the available capacity slack is small. We further find that optimal centralization improves efficiency in all Boroughs, but the gain from centralization is small compared to the equity-efficiency trade-off when neighborhoods are dissimilar.

\paragraph{(b) Simulation optimization framework to design SLAs.} The stylized model -- though tractable for qualitative insights -- excludes several real-life components that prevent its use to optimize an actual city agency's budget. These components include non-Poisson incident arrival rates that exceed available service capacity. We thus develop a simulation-optimization framework to optimize decisions in practice and evaluate the conceptual predictions in the simulation: the simulation inputs historical incident arrivals and worker capacities over time, simulates daily decisions according to a given policy (including per-area budget fractions and incident priorities), and calculates efficiency and equity metrics. Then, we can optimize over a class of policies in an outer loop, such as through a Bayesian optimization framework. Optimal policies can be evaluated out of sample, such as by simulating their performance using data from a future time.  The framework is thus flexible enough to incorporate high-fidelity simulations, given the policy.

\paragraph{(c) Data-driven characterization of SLAs under different objectives.} We apply our simulation-optimization framework to data from the Department of Parks and Recreation in New York City, for responses to service requests made by residents, testing our theoretical predictions and analyzing SLAs in practice. We first find that optimized Borough-budget policies differ from the status quo, which is both inefficient and inequitable. We further find that optimal policies are robust over time: optimal policies calculated using 2019 data are highly effective (outperforming actual decisions, in terms of both efficiency and equity) in future years. Finally, we illustrate through data answers to our two conceptual questions, reflecting the theoretical predictions. (i) We find that, while there is a large efficiency-equity tradeoff at the extremes, large equity gains can be had for relatively low efficiency loss. Furthermore, adding explicit equity terms to the objective has a small impact on efficiency compared with designing better (even purely for efficiency) budget allocations. We explain this finding as follows: inequity often implies inefficiency -- substantial delays in one neighborhood also affect the overall city welfare, and so the equity-efficiency Pareto frontier has many points that perform well on both metrics. This finding suggests a ``win-win'' when compared to the status quo: we can improve both city-level outcomes and equity. (ii) Centralizing operations optimally indeed shifts the Pareto frontier to improve both metrics; however, it only provides a modest benefit over optimal Borough-specific budget allocations, and the gain from centralization is smaller than the price of equity, as predicted theoretically.

Putting things together, our work provides a conceptual framework to reason about efficiency and equity in service level agreements, connecting operational constraints with policy goals. First, we theoretically analyze the equity-efficiency tradeoff and gain from centralization. We then, in a data-driven approach, demonstrate that our modeling insights continue to hold in our calibrated simulations. We further provide a simulation method to produce optimal, robust policies.

This paper is organized as follows. We introduce our model and theoretical results in \Cref{sec:model}, our simulation optimization framework in \Cref{sec:simulationoptimization}, and our NYC Parks application in \Cref{sec:results}. We conclude with a discussion of the communication and iteration of our results with NYC Parks.   

Code and data necessary for replicating our simulation results are available at \url{https://github.com/nikhgarg/sla_design}. Our theoretical statements have been formalized in Lean using the AppliedModelingLib project \cite{garg2026econcslib} and the formalization is available at \url{https://github.com/nikhgarg/AppliedModelingLib/blob/main/papers/LG24ServiceLevelAgreements/}.

\subsection{Related Work}
\paragraph{Efficiency and equity in government operations.} Our application -- responses to service requests made by residents -- relates to previous works on ``co-production'' \citep{yuan_co-production_2019,brabham_crowdsourcing_2015} systems such as the 311 system in NYC, where city residents report incidents, triggering the need for responses. Recent work has studied upstream reporting behavior of city residents, including identifying disparities in usage  \citep{liu2023quantifying, hacker_spatiotemporal_2020, kontokosta_bias_2021, agostini2023bayesian,clark_advanced_2020,o2016311}. 
Recent work has also considered efficiency and equity in public resource allocation. Previous theoretical works on this topic outline tradeoffs among equity criteria and between efficiency and equity \citep{mashiat2022trade, freeman2020best}, in areas such as allocation of healthcare resources \citep{mhasawade2021machine}. Empirical work has considered whether status-quo response decisions are equitable and efficient. Most notably, \citet{singh2022fair} study food inspection operations in Chicago to identify violations of fairness criteria due to idiosyncratic behavior of inspectors, suggesting algorithmic remedies; \citet{rahmattalabi2022learning} and \citet{jo2023fairness} consider resource allocation in homeless services, both identifying incompatibilities among fairness objectives and developing a fair matching algorithm.

In contrast, considering time-sensitive incidents, our work mainly asks the question ``how do we \textit{make} the responses more equitable and efficient by adjusting the agency's resource allocation policy?'' Our work can incorporate insights from the above work: for example, it is possible to incorporate the incident type and location-specific reporting delay estimates by \citet{liu2023quantifying} into our framework -- if the city's objective is to equalize \textit{occurrence to resolution} \textit{end-to-end} response times, as opposed to \textit{report to resolution} response times.

More broadly, our work relates to a growing literature that studies the equity-efficiency trade-off in government operations and allocation of public resources and goods \citep{monachou2022fairness}, both theoretically and in practice, particularly from an operational perspective: e.g., medical supplies \citep{manshadi2021fair}, food donations \citep{sinclair2022sequential, banerjee2023online}, social services \citep{koenecke2023popular,azizi2018designing}, education \citep{Meier2023Effectiveness}, transportation \citep{maheshwari2024congestion, torrico2024equitable, ostrovsky2024effective,jalota2021efficiency,delarue2024algorithmic,banerjee2019incorporating,bertsimas2020bus}, recommendations and ranking \citep{chen2023interpolating,greenwood2024user,manshadi2023redesigning}, and efficiency-equity tradeoffs more broadly \citep{bertsimas2011price,bertsimas2012efficiency}. Our setting fits the framework of \citet{bertsimas2012efficiency}; we analyze the efficiency-equity trade-off for a constant number of agents (Boroughs), under a notion similar to max-min fairness, and study the impact of this trade-off on policy-making. To this line of work, we contribute a computationally tractable approach to finding the {equity-efficiency} Pareto frontier; conceptually, we characterize when this tradeoff may not be large in practice.

\paragraph{Design and planning of queueing systems in operations.} Our problem broadly falls into the ``capacity and flow assignment'' category under the four types of optimization problems tied with the design and planning of queueing systems outlined by \citet{kleinrock1975queue}. Other works that fall into this category include a large literature on dynamic capacity allocation (e.g., \cite{andradottir2003dynamic}) and, in particular, on the design of hospital operations (e.g., \cite{bekker2010time, green2006queueing, cochran2009multi}). More closely aligned with our objectives are the works of \citet{nowak2004bandwidth} and \citet{remesh2019service}; both consider service-level-agreement-aware dynamic capacity allocation in the network service context. Our work differs from these settings in that the capacity allocation decisions cannot be frequently adjusted in our setting -- once the city determines a set of budgets for each Borough, it might take years to revise such decisions, both due to administrative capacity and the challenge of moving workers to different offices.

Our model builds off of the work of \citet{liu2001maximizing}, who consider maximizing profit under SLA constraints; they consider the levers of \textit{prioritizing} and \textit{routing} different jobs to different servers, under the Generalized Processor Sharing \citep{parekh1993generalized} method. Our work introduces a capacity allocation lever, but does not consider routing decisions; these changes, besides being motivated by our application domain questions, render the optimization problem more tractable, allowing us to analytically characterize the price of equity under such a model. We further embed this model within a data-driven simulation-optimization framework, finding approximately optimal policies in practice. This step is outside the scope of the aforementioned works, and represents a feasible path for researchers and policymakers to analyze and recommend policies for practice.

\section{A Stylized Model to Design Optimal Service Level Agreements}
\label{sec:model}

We start with a stylized model of our setting. The model has four aspects under a policymaker's control: how individual workers are allocated to incidents, how many workers there are in each neighborhood (Borough), the SLAs promised for each category (potentially in a neighborhood-dependent manner), and an objective function formalizing their efficiency and equity goals. We further assume that the policymaker knows (can historically measure) characteristics of incidents, such as their arrival rates and average ``riskiness.'' 
These aspects are related as follows. Together, the worker allocation policy, budgets, and incident arrival rates determine the distribution of response times for each incident, and thus the Service Level Agreements. We assume that the policymaker's objective is a function just of the SLAs and the system constants (such as incident arrival rates and risk distributions) -- together, these aspects can encompass standard metrics such as average and tail response times, and their spatial distribution.

First, for a given policy objective, we formulate and solve a corresponding (tractable) optimization problem to find an optimal SLA (and hence a worker allocation and budget policy). Second, we characterize how solutions change with the objective or the ability to centralize. Using this characterization, we answer our research questions. We find that, while efficiency-equity tradeoffs can be large, there are solutions with large equity gains (compared to historical outcomes) with relatively small efficiency loss; furthermore, the gain from centralization may be small in practice. \par

Our theoretical model is purposely stylized to enable tractable insights. In practice, policy design must incorporate elements omitted in the theoretical analysis; we consider these components in \Cref{sec:simulationoptimization} within a simulation optimization framework, and show that these insights hold in data.

\subsection{A Queueing Model for the Inspection Scheduling Process}
We start with a queueing-based model for how incidents are addressed: incidents occur over time, and join a corresponding queue. Workers service each queue according to an allocation policy over time, inducing response time distributions for each type of incident.

\paragraph{The queueing model.} The inspection problem is modeled by a queueing network with multi-class single-server queues. There are two levers of policy: the \textit{city} allocates worker budgets to Boroughs, and each Borough manages the allocation of workers to individual incidents.

The queueing network within each Borough operates as follows. There is a set of incident categories $k \in \mathcal{S}$ (e.g., Hazards versus less urgent incidents) for which we wish to define SLAs. Incidents in each category arrive according to a {Poisson process} with rate $\lambda_{k,b} > 0 $ into their own queue, and these Poisson processes are mutually independent. In practice, not every incident is responded to; we model this as incidents being dropped (independently) with probability $1 - \pi_{k,b} \in [0, 1)$, and define the non-dropped incident rate as $s_{k,b}=\lambda_{k,b}\pi_{k,b}$.
By independent thinning, admitted requests form independent Poisson processes of rates $s_{k,b}>0$.
Inspecting an incident takes a random amount of time, {with i.i.d. unit-mean exponential service times, independent of other events.} \par

The city has a budget of $C$, the total capacity of the servers that can be allocated. The city first decides the capacity $C_b$ to allocate to each Borough $b\in \mathcal{B}$, where $\sum_{b\in \mathcal{B}} C_b\le C$. Each Borough $b$ maintains a server with capacity $C_b$, serving the queues for incidents in that Borough.

We assume that within each Borough, the server is managed with the Generalized Processor Sharing (GPS) scheme \citep{parekh1993generalized}. {We use the ideal-fluid, preemptive-resume version of GPS and evaluate long-run response times after the system has reached steady operation.} Under the GPS scheme, each SLA category is assigned a weight $\phi_{k,b}$ in each Borough $b$, such that $\sum_{k\in \mathcal{S}}\phi_{k,b} \leq 1$. At any given time $t$, on a server with capacity $C_b$, the (potentially fractional)
capacity devoted to category $k \in \mathcal{S}$ is $C_b\phi_{k,b}/\sum_{k'\in \mathcal{K}_b(t)}\phi_{k',b}$, where $\mathcal{K}_b(t)$ denotes the set of categories for which there is at least one pending incident (there is a ``backlog'' of incidents) in Borough $b$. {Thus, every backlogged category $k$ receives service at rate at least $C_b\phi_{k,b}$. }To finish formally describing the system, as standard in such queueing systems \citep{liu2001maximizing}, we assume that there is an additional category of work that is always backlogged for which the city provides no guarantees---the worker services this category when there is no backlog in categories $k \in \mathcal{S}$. Within each queue, we assume a first-come-first-served (FCFS) discipline of service. At a high level, $\phi_{k,b}$ determines the relative priority of different incident categories within a Borough---for example, a \textit{Hazard} category may be higher risk on average, and so prioritized on average.

What are the practical implications of this class of policies? When there are multiple categories of incidents in the backlog in a Borough $b$, each category $k$ of incidents is effectively worked on in parallel queues, by servers with capacity proportional to their assigned weights $\phi_{k,b}$, which reflect their prioritization levels. This can be achieved by assigning teams of inspectors with team sizes proportional to the weights to focus on different categories. When there is only one category of incidents remaining to be inspected, it naturally leads to all inspectors focusing on that remaining category. Within each queue, inspectors go over incidents on an FCFS basis.

Why limit to this class of policies? (a) GPS policies are flexible and robust: 
on a server with capacity $C_b$, the minimum capacity devoted to any SLA category $k$ would be  proportional to $C_b\phi_{k,b}$ while there are incidents in this category in backlog, so the length of the backlog of other SLA categories will not affect our response to category $k$; on the other hand, the maximum possible capacity devoted to any SLA category is the entire capacity $C_b$, whenever it is the only SLA category in the backlog. (b) The policy closely reflects what is practiced by city agencies: a certain portion of workers would respond to certain categories of incidents, with day to day adjustments based on the backlog. Thus, optimizing policies within this class informs us of the realistic gains, without the need to completely overhaul organizational structures. (c) This class of policies and its variants are considered in other fair load balancing applications, e.g., \citet{zhao2019preemptive}. (d) As we will show, GPS policies naturally lead to well-defined notions of SLAs. Under this model, the decision variables are the Borough-level budgets $C_b$ and the incident allocation weights $\phi_{k,b}$.

\paragraph{Response times and SLAs.} Reflecting practice, we consider SLAs in the following form: 
    \textit{``In Borough $b$, at least a fraction {$\frac{s_{k,b}}{\lambda_{k,b}}(1 -\alpha_{k,b})$} of category $k$ {incidents} are responded to by time $z_{k,b}$,''}
where $0<\alpha_{k,b}<1$ and $0<s_{k,b}\leq\lambda_{k,b}$. It is thus important to quantify the tail distributions of the response times of non-dropped incidents. Mathematically, let $T_{k,b}$ be the {random variable corresponding to the response time of non-dropped incidents of category $k\in \mathcal{S}$ in Borough $b$}. Then, an SLA of the above form corresponds to {$\mathbb{P}[T_{k,b} > z_{k,b}]\le \alpha_{k,b}$}. Assuming that these incidents are processed on a server with capacity $C_b$, and the GPS weight of this category satisfies $C_b\phi_{k,b}>{s_{k,b}}$ (i.e., there is guaranteed to be enough capacity to {respond to all admitted requests} of this category), the tail probability of the response time distribution is bounded by
\begin{equation}
\begin{aligned}
    {\mathbb{P}[T_{k,b}> z_{k,b}]}
    &\le \exp\left(-(\phi_{k,b} C_b-{s_{k,b}})z_{k,b}\right).
\end{aligned}
    \label{tail_prob}
\end{equation}

This follows from our assumptions and classical results in queueing theory (e.g., see \cite{kleinrock1975queue} and \cite{liu2001maximizing}). Combining the above, a sufficient condition to satisfy an SLA is
\begin{equation}
    (\phi_{k,b} C_b-{s_{k,b}})z_{k,b}
    \ +\ \log \alpha_{k,b}\geq0,
    \label{def:sla}
\end{equation}
  which we will henceforth refer to as the ``SLA constraint.'' Intuitively, this constraint states that when $C_b$ budget is allocated to Borough $b$, and category $k$ is assigned a GPS parameter $\phi_{k,b}$, then at least a fraction $\frac{s_{k,b}}{\lambda_{k,b}}(1-\alpha_{k,b})$ of all category $k$ requests in Borough $b$ should be {responded to} within $z_{k,b}$ days, i.e., the SLA would be $z_{k,b}$ days. Practically, enforcing the same tail constraint for all categories enhances the interpretability of the policy;
thus, we will consider {$\alpha_{k,b}=\bar\alpha$ and write $a=-\log(\bar\alpha)$} for all category--Borough cells $(k,b)$, which also simplifies notation. Note that {$\bar\alpha\in(0,1)$}, and so {$a>0$}. In the appendix, we prove some of the statements using general $a_{k,b}$. \par

\paragraph{Policymaker's objective.} From the perspective of the city, the objective broadly contains two parts. First, SLAs across the whole city should reflect the principle of efficiency, in that more urgent incidents should be addressed sooner: if a more hazardous incident is left unattended for one day, the amount of risk it poses to surrounding residents is larger than a less hazardous one. However, since responding to service requests is a \textit{public} service, the city must treat different people equitably: residents from different areas should not receive dramatically different levels of service. 

We define two functions to capture these notions. Let $\textbf{z} = \{z_{k,b}: k\in \mathcal{S}, b\in \mathcal{B}\}$ be the set of SLAs across all Boroughs and categories. Denote $g(\textbf{z})$ and $f(\textbf{z})$ by the efficiency loss function and the equity loss function, respectively, meaning that higher values of these functions represent less efficient and less equitable SLAs. We assume that {both functions are} convex in $\textbf{z}$. {The efficiency loss is coordinatewise nondecreasing in the SLA thresholds, but a within-category geographic range used for equity need not be (i.e., improving one SLA may worsen equity metrics).} Within our stylized model, these assumptions lead to tractable analyses, and we present calibrated simulation results with different specifications, both in the main text and in \Cref{app:altobj}.

\subsection{An Optimization Problem for Optimal SLAs}
\label{subsec:optimizationmodel}
Here, we consider optimizing SLAs given the above model. Crucially, in the theoretical analysis, we hold $\pi_{k,b}$ and hence the arrival rate of eventually inspected incidents, $s_{k,b}$, fixed. As discussed below, we do so for theoretical tractability, and relax this in the simulation-optimization analysis.

The optimization task is as follows. {Fix the admitted rates $s$ and any convex loss $L(z)$. }The objective \eqref{opt-obj} reflects efficiency and equity losses, as functions of the SLAs $\textbf{z}$; constraint \eqref{opt-sla} ensures that the solution meets the set of SLAs encoded in $\textbf{z}$, where {$a$ comes} from our definition of SLAs in Eq.~\eqref{def:sla}; constraint \eqref{opt-gps} comes from the GPS scheduling scheme; constraint \eqref{opt-budget} enforces the overall budget constraint across Boroughs.  \begin{subequations}
\begin{align}
    \min_{\textbf{z}\,{>0}, \phi\ge 0, C_b\ge 0} \quad
    & {L(\textbf{z})} \label{opt-obj}\\
    \ \text{s.t.}\quad
    & -(C_b\phi_{k,b} - {s_{k,b}}) z_{k,b}
      + {a} \le 0,\nonumber\\
    & \quad \quad \quad \quad \forall k\in \mathcal{S}, b\in \mathcal{B}, \label{opt-sla}\\
    & \sum_{k\in \mathcal{S}} \phi_{k,b} \le 1, \forall b\in\mathcal{B}, \label{opt-gps}\\
    & \sum_{b\in \mathcal{B}} C_b \le C.\label{opt-budget}
\end{align}
\label{opt-nonconvex}
\end{subequations}
{Problem \eqref{opt-nonconvex} is a convex optimization problem.  For
the analysis below, it is useful to
eliminate the auxiliary Borough capacities and GPS weights.} For the fixed admitted rates $s$, let $E(s)$ be the capacity in excess of admitted load,
  ${E(s)=C-\sum_{k,b}s_{k,b}>0.}$
With unit-mean service times, $\sum_{k,b}s_{k,b}$ is the mean
admitted workload, so capacity slack $E(s)$ is the remaining service-rate slack available to tighten SLA guarantees.
Then, we get an equivalent, tail-threshold-only problem.
\begin{restatable}{prop}{reformulation}[Equivalent reciprocal-capacity representation]
\label{prop:opt-reformulation}
{Fix $s$ with $E(s)>0$ and a convex loss $L$. The fixed-load design
problem is equivalent to the tail-threshold-only problem}
\begin{equation}
  \min_{z>0}\ L(z)
  \quad\textnormal{subject to}\quad
  \sum_{k,b}\frac{a}{z_{k,b}}\leq E(s).
  \label{opt-reformulated}
\end{equation}
\end{restatable}
We use this representation in the analysis below. In \Cref{app:fixed-load-theory-proofs}, we detail the transformation, the construction of supporting Borough capacities and GPS
weights, and establish uniqueness when the reciprocal-capacity constraint is binding. \par

\textit{Model discussion.} What does this model capture? Namely, four aspects under a policymaker's control. It captures how individual workers are assigned to each incident, through the GPS scheme and the decision variables $\phi$; it captures how many workers are allocated to each Borough, through the capacity of the Borough servers using decision variables $C_b$; it captures the SLAs promised for each category and Borough through decision variables $z_{k,b}$; and finally, the objective function is open to configuration for policymakers to reflect their efficiency and equity goals.

This model is also stylized in several crucial respects that render it inappropriate to use to design \textit{actual} government policies as-is; notably, here, we hold fixed $\pi_{k,b}$, the fraction of incidents inspected, and so the admitted rates $s_{k,b}$. We discuss more of these aspects in detail in \Cref{sec:whysimopt}, when detailing our simulation-optimization framework. In this section, we use the stylized model to draw insights on the structure of optimal policies, and equity-efficiency trade-offs. We consider two administrative policies: first, the status quo administrative policy under which the city first allocates budgets to Boroughs, and then Boroughs manage their responses; and second, an approach in which the city manages a centralized server. \par

\subsection{Efficiency-equity Tradeoff and the Gain from Centralization}
\label{subsec:objectives}
{The representation in \eqref{opt-reformulated} yields a
tractable optimization problem for determining SLAs, budgets, and allocation
policies.} We now characterize the solutions of this optimization problem, to understand efficiency and equity tradeoffs and the gain from centralization in our allocation problem.

\subsubsection{Analysis of the Optimization Model under Specific Objectives. }

We start by specifying the efficiency and equity objectives and relative weightings used in our conceptual analysis. 

\paragraph{{Priority-weight-based} objective functions.} An important aspect of an incident is the \textit{risk} it poses if unaddressed (for example, the danger posed by a tree falling on a person or power line).
We assume that we exogenously have a {positive ex ante priority weight} $r_{k,b}>0$ for incidents of category $k$ in Borough $b$. (In practice, incidents within a type may have different true risk ratings; however, the agency may not know the true risk until after the inspection. Since we are modeling inspection decisions themselves, we assume a fixed priority weight per incident category in each Borough that could arise, e.g., from risk averages.)

Motivated by the importance of potential heterogeneity of {priority} across incident types and Boroughs, we {define the following {risk-based cost function} for each Borough, which is a function of these priority scores, {delay costs for inspected incidents}, and a fixed cost per dropped incident. Let $D_{k,b}\geq0$ be the delay-equivalent penalty for a request that is not responded to, and define the cost of a category-$k$ request in Borough $b$ as}
\begin{equation}
    \begin{aligned}
    {\operatorname{Cost}_{k,b}(s,z)}
    &{=r_{k,b}\left[\frac{s_{k,b}}{\lambda_{k,b}}z_{k,b}
      +\left(1-\frac{s_{k,b}}{\lambda_{k,b}}\right)D_{k,b}\right].}
    \end{aligned}
    \label{eq:all-request-burden}
\end{equation}

{The first term is the {delay penalty} for {inspected} incidents, and the {second} term is the cost penalty for not {inspecting} an incident. We define efficiency by total arrival-weighted burden and geographic equity by comparing the same incident category across Boroughs:}
\begin{align}
    G(s,z)&=\sum_{k,b}\lambda_{k,b}\operatorname{Cost}_{k,b}(s,z),
    \label{eq:fixed-load-efficiency}\\
    F(s,z)&=\sum_k\left[\max_b\operatorname{Cost}_{k,b}(s,z)\right.\notag\\[-2pt]
    &\hspace{24mm}\left.-\min_b\operatorname{Cost}_{k,b}(s,z)\right]
    \label{theoretical_equity_loss}
\end{align}
This equity loss asks whether similar requests receive comparable service across Boroughs. Notably, our equity function $F$ is not weighted by incident arrival rates, unlike overall cost $G$ -- in practice, Boroughs differ substantially in the number of incidents, and equalizing overall induced cost is impractical; rather, we equalize {per-request burden}. To trace the trade-off, define
\begin{equation}
    {L_\gamma(s,z)=\gamma G(s,z)+(1-\gamma)F(s,z),}
    \quad \gamma\in[0,1].
    \label{eq:tradeoff-objective}
\end{equation}
Here, we assume {priority weights $r$ and nonresponse penalties $D$} are given as data, and $\gamma\in[0,1]$ is a hyperparameter for the relative importance of the objectives: the larger $\gamma$ is, the more weight is put on efficiency. In words, {$G(s,z)$ represents total all-request burden as the loss of efficiency, and $F(s,z)$ measures within-category geographic disparity as the loss of equity}. Both {$G$ and $F$ are convex; $G$ is coordinatewise nondecreasing in $z$, whereas $F$ need not be}. These formalizations correspond to common notions of efficiency and equity.

For our theoretical results, it will be convenient to define a benchmark-specific effective-load index for the
efficiency-maximizing solution,
$A_{\mathrm{eff}}(s) =\sum_{k,b}\sqrt{a s_{k,b}r_{k,b}}$.\footnote{Any reduced problem of the form
$\min_x\sum_i c_i x_i$ subject to $\sum_i d_i/x_i\leq E$ has optimal variable
cost $(\sum_i\sqrt{c_i d_i})^2/E$, and so $A_{\mathrm{eff}}$ captures the
cost component corresponding to inspected incidents.}

\subsubsection*{Efficiency and equity optimizing solutions.}

We now analyze the \textit{efficiency-equity} tradeoff in allocation, starting by characterizing the optimal solutions under each objective.

\begin{restatable}{prop}{propextremeefficiency}[Extreme efficiency]
\label{prop:extremeeff}
When $\gamma=1$, the optimal efficiency maximizing solution to Problem \eqref{opt-reformulated} {is given by}
\begin{equation}
    {z^{\mathrm{eff}}_{k,b}
      =\frac{{A_{\mathrm{eff}}(s)}}{E(s)}
      \sqrt{\frac{a}{s_{k,b}r_{k,b}}}.}
      \label{eq:fixed-load-efficient-endpoint}
\end{equation}
\end{restatable}

{Recall that $a=-\log(\bar\alpha)>0$, where $\bar\alpha$ is the common
response-time tail tolerance; larger $a$ therefore represents a stricter
reliability requirement.}
In other words, the optimal solution when we only care about efficiency is such that each type of incident will be assigned an SLA that is inversely proportional to the square root of its {inspected priority-weighted load}: more urgent incidents should be {responded to} sooner, holding admitted volume fixed. Interestingly, note that \textit{budgets} may differ substantially across Boroughs, but per-incident service guarantees remain a function of {admitted load and priority}. Now we consider the extreme equity solution.

\begin{restatable}{prop}{propextremeequity}[Extreme equity]
\label{prop:extremefair}
When $\gamma=0$, {the efficiency-best equitable solution $z^{\mathrm{eq}}$ satisfies}
\begin{equation}
    {\operatorname{Cost}_{k,b}(s,z^{\mathrm{eq}})=M_k,
      \qquad k\in\mathcal S,\ b\in\mathcal B.}
\end{equation}
\end{restatable}

In other words, the zero equity loss solution is feasible, and {all Boroughs would experience the same per-incident cost} {for each incident category} at that solution (note that this solution may produce some unnecessarily high delays $z_{k,b}$ in order to equalize costs).

What are the practical implications of these two extremes? In the extreme efficiency case, all SLA categories are assigned SLAs that are {determined by their admitted load and priority}: the higher the {priority weight}, the sooner the incidents are promised to be {inspected}. However, in practice, each Borough has different geographical characteristics and thus different kinds of potential incidents---i.e., {priority weights and arrival rates} differ by area. Suppose all incidents in one Borough are {uniformly riskier} than those in another Borough; then, only optimizing for efficiency may result in a large divide in the level of service that the two Boroughs receive: one Borough {may} receive worse {priority-weighted} service. An equity term balances the differences: in the extreme equity case, there is a strict parity among the {within-category, per-incident costs across Boroughs}. 
We formalize this discussion next in \Cref{sec:equity_efficiency_tradeoff}, characterizing the ``efficiency cost'' of pursuing equity. Additionally, in \Cref{sec:efficiency_gain_from_centralization}, we discuss implications for efficiency when resources are pooled under a centralized, city-budget inspection structure.

\subsubsection{Equity-efficiency Trade-off. }
\label{sec:equity_efficiency_tradeoff}

When is the equity-efficiency trade-off substantial? Define the \textit{price of equity} as the difference in efficiency loss between {the most equitable and most efficient solutions}, relative to the efficiency loss of the most efficient solution:
\begin{equation}
{\operatorname{PoE}_C(s)
:=\frac{G(s,z^{\mathrm{eq}})- G(s,z^{\mathrm{eff}})}{G(s,z^{\mathrm{eff}})} \geq 0.}
\label{eq:price-of-equity-definition}
\end{equation}
{We refer to the unnormalized difference $G(s,z^{\mathrm{eq}})-G(s,z^{\mathrm{eff}})$ as the (absolute) price of equity.}

For either endpoint policy $z^j$ for $j\in\{\mathrm{eff},\mathrm{eq}\}$, let $q^j=q(z^j)$ denote the fraction of total capacity slack $E(s)$ assigned to category--Borough cell $(k,b)$, where
$q_{k,b}(z^j)=\frac{a}{E(s)z^j_{k,b}}$.\footnote{The supporting policy satisfies $\phi_{k,b}C_b=s_{k,b}+a/z_{k,b}$, so $a/z_{k,b}$ is the cell's guaranteed service-rate margin above its admitted load.  Because the reciprocal-capacity constraint binds, these margins sum to $E(s)$; hence
$q_{k,b}(z)=(a/z_{k,b})/E(s)$.} For probability vectors $P$ and $Q$, let the Pearson chi-square divergence be $\chi^2(P\|Q)=\sum_{k,b}(P_{k,b}-Q_{k,b})^2/Q_{k,b}$.

\begin{restatable}{prop}{fixedloadalignment}[Price of equity]
\label{prop:costofequity}
The price of equity satisfies
\begin{align}
\operatorname{PoE}_C(s)
&=\frac{A_{\mathrm{eff}}(s)^2}{E(s)G(s,z^{\mathrm{eff}})}
 \chi^2\!\left(q^{\mathrm{eff}}\middle\|q^{\mathrm{eq}}\right)
 \label{eq:price-of-equity-chi-square}\\[-2pt]
&\leq \frac{1}{G(s,z^{\mathrm{eff}})}\sum_{k,b}\lambda_{k,b}\notag\\[-2pt]
&\quad\times\left[
 \begin{gathered}
 \max_{b'\in\mathcal B}\operatorname{Cost}_{k,b'}(s,z^{\mathrm{eff}})\\[-2pt]
 {}-\operatorname{Cost}_{k,b}(s,z^{\mathrm{eff}})
 \end{gathered}\right].\notag
\end{align}
The price is zero if and only if
$q^{\mathrm{eq}}=q^{\mathrm{eff}}$, equivalently if and only if
$\operatorname{Cost}_{k,b}(s,z^{\mathrm{eff}})$ is constant across Boroughs for each  category $k$.  Moreover, suppose the fixed noninspection burden
$r_{k,b}D_{k,b}\left(1-s_{k,b}/\lambda_{k,b}\right)$ is constant across
Boroughs within each category.
Then, writing $z^{\mathrm{eff}}(C)$ for the efficient solution at capacity $C$,
\begin{equation}
\operatorname{PoE}_{C'}(s)
=\frac{C-\sum_{k,b}s_{k,b}}
       {C'-\sum_{k,b}s_{k,b}}
 \frac{G(s,z^{\mathrm{eff}}(C))}{G(s,z^{\mathrm{eff}}(C'))}
 \operatorname{PoE}_{C}(s)
\label{eq:poe-capacity-scaling}
\end{equation}
for all $C'>C>\sum_{k,b}s_{k,b}$.
The multiplicative factor in Eq.~\eqref{eq:poe-capacity-scaling} is less than one whenever
$\sum_{k,b}(\lambda_{k,b}-s_{k,b})r_{k,b}D_{k,b}>0$; it equals one when this fixed noninspection component is zero.
\end{restatable}
If primitives $\lambda_{k,b}$, $s_{k,b}$, $r_{k,b}$, and
$D_{k,b}$ are identical across Boroughs within every category, the price is
zero at every capacity; more generally, the price of equity depends on how different Boroughs are in the rates and risks of incidents that occur. Furthermore, the relative price of equity decreases with capacity when the (exogenous) noninspection component is positive. ({Note that full admission, $s=\lambda$, as may occur in practice with substantial capacity, induces all noninspection burdens to be zero, and thus equal.})

Intuitively, the \textit{price of equity} {in absolute terms} incurred by pursuing the most equitable solution compared with the most efficient solution is always non-negative and depends on two factors\footnote{For this discussion, we set aside the choice of incident drop rates $(1 - \pi_{k,b})$, and assume that per-incident noninspection burdens are equal across Boroughs within a category. In the simulation, we show that the insights continue to hold when $\pi$ is endogenous.}: (1) differences in incident occurrence rates across Boroughs within a category; and (2) the capacity slack $E(s)$, i.e., the additional capacity on top of the rate of addressed incidents.

When is the trade-off negligible {in absolute terms}? (1) When {$\operatorname{Cost}_{k,b}(s,z^{\mathrm{eff}})$ is constant across Boroughs within every category}, the price of equity is zero. The intuition behind this is that, when the Boroughs have the same incident profile, they are essentially indistinguishable in the allocation problem. As in other work in the algorithmic fairness literature (e.g., \cite{cohen2022price}), there is no trade-off between equity and efficiency---the objectives are perfectly aligned with homogeneous groups. (2) When {capacity slack $E(s)$} is large---i.e., with enough budget to address all incidents quickly---there is little tradeoff in absolute terms.
In contrast, when is the trade-off substantial {in absolute terms}? We see that two conditions make the trade-off substantial: (1) when Boroughs face sufficiently different request situations, and (2) when capacity slack $E(s)$ is sufficiently small. This is also analogous to the equity-efficiency trade-off in other algorithmic decision-making settings. For example, in college admissions (e.g., \cite{garg2020dropping}), if the admission capacity of a school is larger than the applicant pool, then admitting everyone would be both efficient and equitable; it is only when the capacity is smaller than the applicant pool and different groups of applicants are disparate that we observe a significant equity-efficiency trade-off. In our setting, when capacity is large, {at both optimized endpoints,} incidents in all Boroughs can be inspected within a reasonable time (as can be observed from the formula for $z^{eq}$ and $z^{\mathrm{eff}}$); when capacity is scarce or noninspection burdens differ across Boroughs, we expect a large trade-off. \par

In NYC, we expect both conditions for a substantial trade-off {in absolute terms} to hold, and thus the (absolute) price of equity to be non-negligible, although the magnitude remains an empirical question. Differences in Boroughs are significant: the number of service requests originating in each Borough varies substantially, as shown in App. \Cref{tab:number_srs_by_borough_year}, which serves as a proxy for the incident arrival rates $\lambda_{k,b}$. Moreover, agencies typically have very little, if any, excess budget on hand. In the case of NYC Parks, a substantial share of incidents is not eventually responded to in the data year studied. However, it is important to note that the theoretical analysis is about the extreme endpoints of maximizing efficiency or equity; there may be intermediary policies with a small loss on each.

\subsubsection{Efficiency Gain from Centralization. }
\label{sec:efficiency_gain_from_centralization}

So far, we have analyzed a model in which agencies allocate Borough-specific budgets that cannot be cross-dispatched in regular operations. NYC Parks is considering \textit{centralizing} efforts: instead of apportioning budget and personnel to each Borough, the city maintains a centralized workforce (budget), and responds to incidents according to their category $k$, regardless of the Borough $b$ they are from. The decision variables are thus reduced to one weight parameter $\phi_k$ and one SLA $z_k$ per category $k\in \mathcal{S}$. 
Writing $G^{\mathrm{city}}(s,z)$ for $G(s,z)$ evaluated under the Borough-invariant thresholds $z_{k,b}=z_k$, the city-budget efficiency problem is
\begin{equation}
{\min_{z>0}}\quad G^{\mathrm{city}}(s,z)
\quad {\textnormal{s.t.}}\quad
{\sum_k\frac{a}{z_k}\leq E(s).}
\label{opt-centralized}
\end{equation}
Let $z^{\mathrm{city}}$ solve Problem~\eqref{opt-centralized}, and define
${A^{\mathrm{city}}(s)}=\sum_k\sqrt{a\sum_b s_{k,b}r_{k,b}},$
analogously to above. Then, the relative centralization gain is
\begin{equation}
\frac{G(s,z^{\mathrm{eff}})-G^{\mathrm{city}}(s,z^{\mathrm{city}})}{G(s,z^{\mathrm{eff}})}
=\frac{A_{\mathrm{eff}}(s)^2-A^{\mathrm{city}}(s)^2}
{E(s)G(s,z^{\mathrm{eff}})}.
  \label{eq:aligned-pooling-gain}
\end{equation}

(With an analogous extreme equity prioritization formulation under city budget, an optimal solution of the form in \Cref{prop:extremefair} may not always be attained, because one category-level threshold $z_k$ need not equalize $\operatorname{Cost}_{k,b}$. Thus, we mainly focus on the extreme efficiency prioritization formulation here.) Denote the optimal solution to Problem~\eqref{opt-centralized} by $\textbf{z}^{city}=\{z^{city}_k: k\in \mathcal{S}\}$, representing the SLAs for each category of incidents across the city. The next result shows that in this extreme efficiency prioritization case, this centralized response indeed provides improvements across \textit{all} Boroughs and categories of incidents, compared to the Borough-efficient benchmark.

\begin{restatable}{prop}{propcentralized}[Centralization gain]
\label{prop:strict_improvements_from_centralization}
Assume $|\mathcal B|\geq2$.
For any $k\in \mathcal{S}, b\in \mathcal{B}$, the inspection-delay SLA under the extreme efficiency prioritizing city-budget policy is strictly better (shorter) than under the extreme efficiency prioritizing Borough-budget policy:
\begin{equation}
z^{city}_{k} < z^{\mathrm{eff}}_{k,b}.
\label{eq:pooling-sla-dominance}
\end{equation}
Furthermore, the centralization gain in efficiency can be compared to the price of equity as follows:
\begin{equation}
\begin{aligned}
  &{\textnormal{centralization gain}
    \geq\textnormal{price of equity}}\\[-2pt]
  &{\qquad\Longleftrightarrow\qquad
    1-\frac{A^{\mathrm{city}}(s)^2}{A_{\mathrm{eff}}(s)^2}
    \geq\chi^2\!\left(
    q^{\mathrm{eff}}\middle\|q^{\mathrm{eq}}\right).}
\end{aligned}
\end{equation}
\end{restatable}

By pooling resources, all Boroughs are better off than under the most efficient Borough-level allocation. We can further characterize the scale of such improvements, comparing the efficiency loss of the equity-maximizing solution with the efficiency gain of centralizing.

As discussed above, the right-hand side term is small when Boroughs are similar in their incident profiles, and can be large otherwise. The left-hand side is small when ${A^{\mathrm{city}}(s)} \approxeq A^{\mathrm{eff}}(s)$; equivalently, when $\sum_k\sqrt{a\sum_b s_{k,b}r_{k,b}} \approxeq \sum_{k,b}\sqrt{a s_{k,b}r_{k,b}}$. This occurs when one Borough $b$ dominates the sum $\sum_b s_{k,b}r_{k,b}$, i.e., when Boroughs are highly dissimilar within each category. Thus, we expect the centralization gain to be larger when Boroughs are similar, and smaller than the price of equity when they are dissimilar. In NYC, we expect Boroughs to be dissimilar, and so the centralization gain to be smaller. This is especially relevant for policymakers: instead of pooling resources, managing current inspection structures while optimizing for better budget allocation could be almost as effective, though there is a gain from centralization. (Our approach does not capture other potential gains from centralization, such as more efficient centralized management structures, emergency response, and pooling of other resources, such as for oversight and administration. Furthermore, the comparison depends on model parameters, suggesting the importance of data-driven simulations.)

Altogether, our conceptual framework provides two predictions for our calibrated simulations: (1) At the extremes, the equity-efficiency trade-off is non-negligible in practice, driven by differences among Boroughs and capacity scarcity. (2) Centralizing budgets is beneficial to all Boroughs, but the gain from centralization is small compared to the price of equity (at the extremes). In \Cref{sec:results}, we provide evidence that these two predictions hold. We further show that, outside the extremes, there exist policies that provide both efficiency and equity gains over the status quo.

\FloatBarrier

\section{Simulation Optimization Framework}
\label{sec:simulationoptimization}

Above, we formulate a theoretically tractable policy optimization problem, whose solutions yield insights regarding the tradeoffs between efficiency and equity in our temporal allocation setting. However, the \textit{optimal policies} are not generally \textit{deployable} -- such theoretical tractability requires omitting modeling components that are important in practice. Thus, an important question is whether the theoretical insights continue to hold with more realistic policies. 

We advocate a simulation-optimization approach for such policy optimization and analysis: (a) specify inputs (here: incident arrivals and city-wide capacities), either by calibrating a model to historical data or by using the data directly; (b) determine a class of policies over which one will optimize (e.g., GPS policies with Borough budgets, incident prioritization, etc.); (c) simulate a given policy over time and evaluate its performance; (d) optimize over policies; (e) finally, evaluate the chosen optimal policy in our calibrated simulations out-of-sample (such as using input data from another historical time period) to understand its robustness and to approximate future performance.

This procedure is of course more computationally expensive than solving the convex optimization Problem \eqref{opt-reformulated}, which {takes less than a second} to solve (as opposed to the months of CPU time required in our NYC Parks application). However, policies are not updated often (e.g., agencies often plan budgets on a yearly cadence) and simulations can be parallelized, and so the cost is not prohibitive. Such an approach can further trade off optimization tractability against modeling fidelity, by adjusting the complexity of the simulator and class of policies.

In this section, we provide methodological details of this procedure for optimizing NYC agency service allocation policies. The next section provides application data details and results.

\subsection{Why a Simulation Optimization Framework}
\label{sec:whysimopt}

Our theoretical optimization framework makes at least two simplifications that prevent its use in practice in our application: (a) Historically, not all incidents are inspected: the number of service requests received far exceeds the capacity of the city agency (in our application, historical data show that a substantial share of requests is not inspected); (b) as seen in \Cref{fig:srs_ins}, incidents do not arrive according to a homogeneous Poisson arrival rate: there are seasonal effects, emergency storm periods, and daily variations not necessarily explainable by Poisson randomness. As we explain next, these components prohibit tractable (convex) policy optimization.

\paragraph{Over-capacity queues and dropping incidents.}

Not all incidents, even in daily operating periods (outside of emergency periods), are inspected by the agency, let alone fixed through a work order. As expected, the fraction of incidents dropped correlates with incident importance (most Hazard reports are inspected, but few Root/Sewer/Sidewalk reports are); however, incident drops can also be an undesirable implication of historical policies: e.g., if a Borough is relatively understaffed, then it may be forced to drop more incidents. Thus, we need to make the fraction of incidents inspected a variable, either optimized directly as part of the policy or determined by capacities and prioritization. However, it turns out that doing so while maintaining theoretical tractability is challenging. Let {$\pi_{k,b}$} denote the fraction of service requests inspected.

First, suppose we fix constants $\pi_{k,b}$ and the total budget $C$ as given and estimated from the historical data, i.e., substitute $\lambda_{k,b}$ in Problem \eqref{opt-nonconvex} with $\pi_{k,b}\lambda_{k,b}$. Then, we would find that if $\sum_{b\in\mathcal{B}}\sum_{k\in\mathcal{S}}\pi_{k,b}\lambda_{k,b} = C$, i.e., the admitted load exhausts the total budget, then any allocation with nonnegative service-rate margins must have $C_b\phi_{k,b}-\pi_{k,b}\lambda_{k,b} = 0$ for all $(k,b)$ (i.e., the arrival rate and the service rate of the queue are the same), and thus no finite SLAs $z_{k,b}$ are feasible.

Second, suppose that $\pi_{k,b}$ is a decision variable, that the objective contains a term $h(\boldsymbol\pi)$ penalizing dropped incidents (where ${\boldsymbol\pi}=\{{\pi_{k,b}}:k\in \mathcal{S}, b\in \mathcal{B}\}$), and that the constraint in \Cref{opt-sla} is imposed:
\begin{equation}
    -(C_b\phi_{k,b} - \pi_{k,b}\lambda_{k,b}) z_{k,b} + a \le 0,  \forall k\in \mathcal{S}, b\in \mathcal{B}. \label{opt-sla-drop}
\end{equation}

The difficulty comes with combining these terms alongside a cost term that penalizes delays for non-dropped incidents, e.g., the bilinear term $\pi_{k,b}z_{k,b}$, which is neither convex nor concave in
$(\pi_{k,b},z_{k,b})$. While it may be possible to re-parameterize the optimization to obtain convexity, it is challenging to do so while keeping objective terms for both delays for non-dropped incidents and an equivalent cost for dropped incidents. Especially due to the additional challenges discussed next, we do not do so theoretically, but instead analyze our model through calibrated simulations informed by a Bayesian optimization approach.

\paragraph{Arrival rates and service times.}
The theoretical analysis further depended crucially on incidents arriving according to a homogeneous Poisson process, and service times (how long it takes a worker to address an incident) being exponentially distributed; these assumptions make it simple to bound the tail probabilities of incident response times, which are part of the SLA. While such assumptions are common in the literature for tractability, they are too strong in our setting for several reasons.

First, incident arrivals do not follow homogeneous Poisson processes, displaying spatial and temporal (weekly and seasonal) correlation, both during emergency periods and in routine operations. Second, incident response times are not necessarily exponentially distributed: for example, in our \textit{inspection} allocation application, an inspector travels to an incident and conducts an inspection, which may take an approximately constant amount of time; some incidents may require follow-up inspections at a later date, which would increase the total time it takes to address that incident but does not preoccupy the inspector's time in the meantime.  Additionally, not all incidents are inspected in a first-come, first-served manner. From 2019 to 2023, around 30\% of inspected Hazard incidents were inspected later than another incident from the same category that got reported later.

Suppose one wanted to incorporate these nuances into Problem \eqref{opt-nonconvex}, using alternate functional forms of arrival and service times, for example incorporating spatial and temporal correlation of arrivals. Theoretically, this would require replacing constraint \eqref{opt-sla} with another constraint that better reflects the mapping from budgets and prioritizations to the tail inspection probabilities that are induced; closed-form solutions may be intractable, depending on the functional forms chosen. For a data-driven analysis, one would need to \textit{calibrate} the functions using real-world data, and high-fidelity calibration may be challenging even if the theoretical challenge could be overcome.

In light of these obstacles, we employ a simulation-based approach to finding near-optimal inspection policies in practice, while retaining the core of our principled approach -- a parametric GPS scheme. We leave theoretical approaches to overcoming these obstacles for future work.

\subsection{Data-Driven Simulation-Optimization Framework}
\label{sec:simpipeline}
\noindent We now describe our simulation-optimization framework, informed by the above discussion. The agency inspects incidents every day, over a 3-year simulated period. As outlined above, we need to specify the input data, policy class, and evaluation.

\subsubsection{Input Data: Historical Inspection Arrivals and Agency Capacity. } We directly use {historical incident arrivals (time stamps, locations, and category)} over a full year (in our application, 2019), with {incidents arriving each day according to their true arrivals}. As inspection budgets may vary throughout the year and week (e.g., fewer inspections on weekends), for each day, we use {the number of inspections the agency performed citywide that day}. (Of course, the policy will determine how this overall daily budget is distributed across Boroughs and incident categories). We repeat the year of data 3 times in each simulation to minimize boundary effects. These choices enable simulation realism without calibration of functions for incident arrivals and capacity.

\subsubsection {Policy Classes. } We optimize over two classes of policies: Borough-budget GPS policies, as in our theoretical model, and city-budget policies that allow more flexible cross-Borough allocation in response to real-time queue lengths.

\paragraph{Borough budget GPS policies.} Each policy consists of two levers, identical to our GPS policy scheme described above: a set of budget allocations $\{C_b\}$ to each Borough that describes the \textit{fraction} of the daily capacity available to that Borough (without loss of generality, we standardize them so that $\sum_{b}C_b = 1$); and two sets of parameters $\{\phi_{k,b}\}$ and $\{p_{k,b}\}$, which indicate how each Borough should manage its inspections under a GPS scheduling scheme, and the probability with which a pending category-$k$ request in Borough $b$ is retained at each scheduled backlog review.

\paragraph{City budget policies.} In addition to evaluating Borough budget policies (where the agency has two levers: budget allocation to Borough subdivisions and managing queues within each Borough), we consider a class of policies that assumes the city agency only manages \textit{one centralized server}, i.e., it can flexibly allocate its workers across Boroughs in reaction to current queues, without needing to worry about e.g., travel time. We refer to this type of policy as \textbf{city budget} policies. City budget policies allow more flexibility than each Borough managing its own server, as when one Borough is experiencing a surge of inspection requests, the capacity that would have been dedicated to other Boroughs can be concentrated on inspecting these. This added flexibility may induce higher administrative and logistical expenses: inspectors are usually stationed within each Borough, and there is a cost to transition them between Boroughs. Under this class, the only lever is how the inspections should be managed, specified by two sets of parameters $\{\phi_{k,b}\}$ and $\{p_{k,b}\}$.\footnote{This approach generalizes the centralization policy analyzed theoretically. The theory pools Boroughs within each category into a single queue; the simulation pools capacity citywide but retains category-Borough-specific queues and controls.}

\noindent These policies are neither exhaustive nor likely to describe exactly how the agency behaves. However, the policy class accurately reflects the agency data and our conversations with agency practitioners. Higher fidelity policies and simulators can be designed (e.g., those that incorporate city budgets but with travel costs for inspectors, spatial grouping of inspected incidents, or more deterministic prioritization of the highest priority incidents), at the cost of an increase in parameters.

\subsubsection{Simulation Given a Policy and Input Data. } \label{sec:sim_given_policy} We now summarize our simulation, with details appearing in \Cref{sec:alt_policy}. There are two hyperparameters for calibrating the simulator to historical inspections: review period length $\{\tau_b, b\in \mathcal{B}\}$ and first-come, first-served (FCFS) violation $\rho$. For a period indexed by $t\in [T]:=\{1,2,\dots, T\}$, the simulation takes in {historical arrivals of requests} ${N}_{k,b}^{t}, \forall k,b$, and {historical citywide inspections performed on day $t$}, denoted $I^t$, and simulates outcomes (whether and when each incident is inspected) under counterfactual policies.

On each day, the new {incidents} join their category--Borough queues and the observed citywide {inspection count} $I^t$ is allocated first across Boroughs according to $\{C_b\}$ and then across nonempty category queues according to $\{\phi_{k,b}\}$. Allocations exceeding a category's backlog are reallocated to other incidents in the Borough if available. Within a queue with backlog $B_{k,b}^t$, the $I_{k,b}^t$ inspections are allocated uniformly at random from the earliest $\rho(B_{k,b}^t- I_{k,b}^t) + I_{k,b}^t$ incidents in the backlog (when $\rho = 1$, it is equivalent to randomly inspecting incidents in the backlog, and when $\rho = 0$, it is strictly FCFS).
As discussed above, historical incident arrival rates are larger than historical budgets can handle, so we specify {per-review retention probabilities} $\{p_{k,b}\}$ as decision variables.
After each period of $\tau_b$ days, each inspection request in the backlog is dropped with probability $1-p_{k,b} \in [0, .9]$ for the appropriate $k$ and $b$. This mimics such a process in practice: when faced with a list of inspection requests, inspectors in each Borough with capacity constraints would only inspect those with perceived urgency above a certain threshold, and such decisions are reviewed on a regular basis. If the incident is not dropped, it stays in its backlog.

This specification of review frequency ($\tau_b$) and FCFS violation ($\rho$) resolves a tension between objectives based on the delay quantile and FCFS, restricted to the same category of incidents: if shorter tail delays are the only objective, then it is desirable to discard all incidents that are past a certain age in the backlog, and maintain stricter FCFS inspections. This, however, overlooks the potential importance of those delayed incidents and the need for more nuanced prioritization within a category. By calibrating hyperparameters $\rho$ and $\tau_b$ using historical outcomes of inspection delays (presented in \Cref{app:supp_data}), we try to capture how Borough managers are balancing these needs.

\subsubsection{Evaluation Metrics. } 
\label{sec:evaluation_metrics}
After each simulation, we take the inspected incidents and calculate the realized SLAs $\hat{z}_{k,b}$ by evaluating the actual median inspection delays for category $k$ in Borough $b$. We further calculate the realized inspection fraction $\widehat{\pi}_{k,b}$ as the fraction of arrivals that are inspected. Both dropped incidents and incidents still in the backlog at the end of the simulation are considered uninspected (and thus each reduces $\widehat{\pi}_{k,b}$). The theoretical model fixes the eventual inspection fractions $\boldsymbol{\pi}$ and bounds the SLAs $\textbf{z}$; through simulation, we evaluate and try to optimize their closest simulation analogues, the realized inspection fractions $\widehat{\boldsymbol{\pi}}$ and median delays $\hat{\textbf{z}}$. Cost, efficiency, and equity are defined analogously to their theoretical counterparts in \Cref{subsec:objectives}:
\begin{align}
\operatorname{Cost}_{k,b}(\hat{\textbf{z}},\widehat{\boldsymbol{\pi}})
&=r_{k,b}\Bigl[\widehat{\pi}_{k,b}\hat{z}_{k,b}
+100(1-\widehat{\pi}_{k,b})\Bigr]\label{eq:cost}\\
G(\hat{\textbf{z}}, \widehat{\boldsymbol{\pi}})
    &= \sum_{k\in\mathcal{S},b\in\mathcal{B}}N_{k,b}
    \operatorname{Cost}_{k,b}(\hat{\textbf{z}},\widehat{\boldsymbol{\pi}})\label{eq:emp_efficiency}\\
F(\hat{\textbf{z}}, \widehat{\boldsymbol{\pi}})
    &= \sum_{k\in \mathcal{S}} \left[
    \max_{b\in \mathcal{B}}\operatorname{Cost}_{k,b}
    -\min_{b\in \mathcal{B}}\operatorname{Cost}_{k,b}
    \right].
    \label{eq:emp_equity}
\end{align}
Here ${N}_{k,b}$ is the number of incidents of category $k$ in Borough $b$ and $r_{k,b}$ is the corresponding category priority weight. The penalty on the uninspected fraction of incidents can be seen as an equivalent cost of 100 days for non-inspection. We analyze sensitivity to this cost in \Cref{fig:pfvarsla}.

\subsubsection{Bayesian Optimization for Near-optimal Policy. } 

Once the simulation framework is defined, finding policies that are (near-)Pareto optimal requires (1) efficient parallelized policy evaluations and (2) efficient iteration of policies in a high-dimensional space without gradient information. State-of-the-art Bayesian optimization techniques in multi-objective optimization provide tools that tackle exactly this task. We use BoTorch \citep{balandat2020botorch}, a framework for Bayesian optimization, and in particular, we use the qNEHVI criterion \citep{daulton2021parallel} {to explore the frontier between the efficiency and equity endpoints; we supplement it with separate single-objective searches targeting those endpoints}.

\subsubsection{Comparison with theory.} Putting things together, the simulation shares objective functions and policy classes with the theory. It differs by endogenizing inspection fractions $\hat \pi_{k,b}$ (through optimization of $p_{k,b}$) instead of exogenously specifying them, using incident arrivals and inspection capacity from data, and allowing for FCFS processing violations. Furthermore, while the theoretical program optimizes $z_{k,b}$, uniquely inducing $\phi_{k,b}$ and $C_b$, in the simulation-optimization framework, we optimize $\phi_{k,b}$ and $C_b$, inducing $\hat z_{k,b}$. One implication of the resulting difference is that zero equity loss may not be feasible in the simulation, as the simulator does not artificially restrict capacity from being used if available (which zero equity loss may require). \par

\FloatBarrier

\section{Data-driven Application}
\label{sec:results}
\noindent We now apply our framework to data from the New York City Department of Parks and Recreation. 

\subsection{Data and Methods Description}
\noindent We use publicly accessible data from the NYC Open Data Portal, which include both inspection requests submitted to the forestry unit of NYC Parks, and inspections performed. Our main policy evaluations and selection use historical data from the calendar year 2019 and out-of-sample evaluations of their performance use historical data from 2021 and 2022. Our 2019 simulation includes {a total of \num{75076} inspection requests} (to be consistent with historical outcomes, we duplicate requests that required additional inspections). We exclude data from 2020 because, as can be observed from \Cref{fig:srs_ins}, requests and inspections for that year are dominated by a tropical storm in August 2020; operational policies during such emergency periods differ substantially from those during regular periods -- substantial cross-agency and cross-Borough resources are allocated to the affected areas, and so optimal policies take a different structure and must be calculated separately.

We consider the task of allocating inspections to reported incidents: in these inspections, a trained forester visits the site and rates the incident on several dimensions, producing an overall risk rating. These risk ratings are then used to help allocate maintenance resources.\footnote{The agency does not have risk ratings for uninspected incidents, and so cannot use incident level risk to allocate inspections; it only has access to the incident category for this decision. We confirm through a regression that, for inspected incidents, risk ratings do not add substantial predictive power for the initial inspection delay, consistent with this information structure.}

Each qNEHVI iteration (which includes optimizing the qNEHVI acquisition function to get a batch of 64 new policies, and then evaluating these policies through the simulation described) {takes between 60 minutes and 24 hours (for later batches) on a modern machine using 32 CPU cores and 128 GB of RAM}. For the endpoint searches, we run 50 batches (and so evaluate 3200 policies); for the intermediate qNEHVI searches, we use 36 batches, as later batches become computationally prohibitive. In \Cref{fig:policyeval}, we present {search progress as a function of} the number of policies evaluated, finding that improvements are small for later batches, implying search convergence.

In the historical data, we find that the average risk ratings of {inspected} incidents are (1) substantially similar across Boroughs for the same category and (2) insufficiently different across categories to reflect the differences in perceived risk by practitioners in our conversations. Thus, we assign {\ category priority weights} to these categories when calculating the efficiency (\Cref{eq:emp_efficiency}) and equity (\Cref{eq:emp_equity}) loss functions, as indicated in App. \Cref{tab:avgrisk} and \Cref{fig:riskdist_hazard}.

\begin{table}[tb]
\centering
\small
\resizebox{.9\textwidth}{!}{\begin{tabular}{clrrr}
\toprule
\begin{tabular}[c]{@{}c@{}}Data source \\ (Calendar year)\end{tabular} &
  \begin{tabular}[c]{@{}c@{}}Objective\\ function\end{tabular} &
  \begin{tabular}[c]{@{}c@{}}Most efficient policy\\ (Relative to historical)\end{tabular} &
  \begin{tabular}[c]{@{}c@{}}Most equitable policy\\ (Relative to historical)\end{tabular} &
  \begin{tabular}[c]{@{}c@{}}Balanced policy\\ (Relative to historical)\end{tabular}\\ \midrule
\multirow{2}{*}{2019} & Efficiency loss  & \textbf{0.800} & 0.902 & 0.808\\
                      & Equity loss  & 0.825  & \textbf{0.426} & 0.447 \\ \midrule
\multirow{2}{*}{2021} & Efficiency loss  & 0.758 & \textbf{0.686} & 0.737 \\
                      & Equity loss  & 0.664  & \textbf{0.457} & 0.521  \\ \midrule
\multirow{2}{*}{2022} & Efficiency loss  & \textbf{0.660} & 0.718 & 0.665 \\
                      & Equity loss  & 0.621  & 0.532 & \textbf{0.495}  \\ \bottomrule
\end{tabular}}
\caption{Performance of the most efficient, most equitable, and Balanced Borough budget GPS policies, when evaluated on data from 2019 (``training set''), 2021, and 2022 (``test sets''), relative to the corresponding efficiency and equity losses calculated from historical inspection data. A lower value indicates better performance, and the best performer in each row is indicated in bold. We omit 2020 due to a large storm dominating the number of incidents; during such emergency periods, the city activates cross-agency resources and does not follow its regular operations. Values are means across 25 simulations for 2019 and five simulations for each out-of-sample year.}
\label{tab:robustnesstab}
\end{table}

\begin{table}[tb]
\centering
\small
\resizebox{.8\textwidth}{!}{\begin{tabular}{lccccc}
\toprule
\multicolumn{1}{c}{\multirow{2}{*}{Borough}} &
  \multirow{2}{*}{\begin{tabular}[c]{@{}c@{}}Fraction of\\ inspection requests\end{tabular}} &
  \multicolumn{4}{c}{Budget allocation} \\ \cmidrule{3-6}
\multicolumn{1}{c}{} &
  & Historical & Most Efficient & Most Equitable & Balanced\\ \midrule
The Bronx            & 0.100 & 0.116       & 0.093           & 0.111        & 0.092   \\
Brooklyn             & 0.299 & 0.295       & 0.322           & 0.342        & 0.310   \\
Manhattan            & 0.073 & 0.072       & 0.075           & 0.082        & 0.091   \\
Queens               & 0.400 & 0.396       & 0.380           & 0.354        & 0.386   \\
Staten Island        & 0.128 & 0.121       & 0.130           & 0.111        & 0.121   \\ \bottomrule
\end{tabular}}
\caption{Comparison of the historical allocation and the allocations of the most efficient, most equitable, and Balanced Borough budget policies with the fraction of inspection requests received in each Borough in 2019.}
\label{tab:budgetalloc}
\end{table}

\begin{figure}[tb]
    \centering
    \includegraphics[width=.95\textwidth]{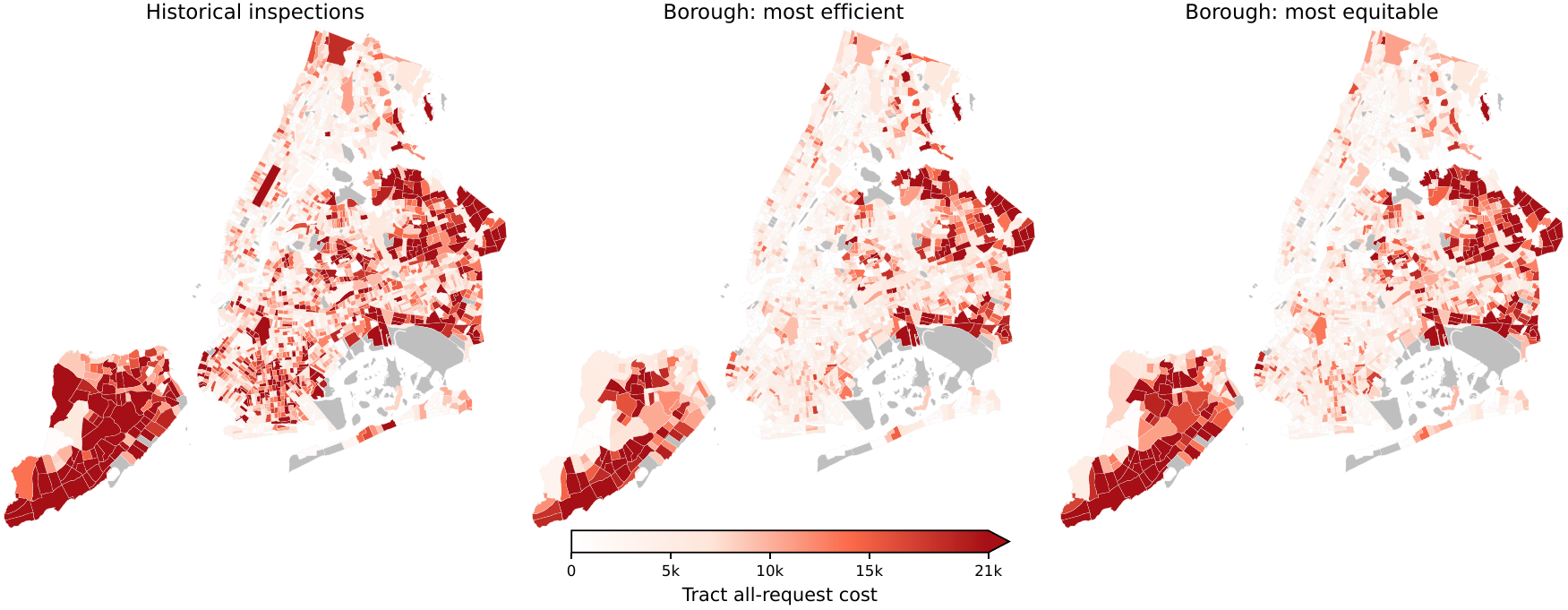}
    \caption{Cost borne by each census tract under three different inspection policies: (a) historical inspections, (b) the most efficient Borough budget policy, and (c) the most equitable Borough budget policy, evaluated on data from 2019. Instead of evaluating $\widehat{\pi}_{k,b}$ (fraction of inspected incidents) and $\hat{z}_{k,b}$ (median of inspection delays) for each Borough $b$, we evaluate them for each census tract in NYC, which are finer-grained subdivisions of Boroughs. We then calculate the total cost for each census tract analogous to \Cref{eq:emp_efficiency}. Note that the three plots are colored using the same scale. The most efficient and equitable policies are far more similar to each other than they are to the status quo.}
    \label{fig:costmap}
\end{figure}

\begin{figure}[tb]
    \centering
    \includegraphics[width = 0.75\textwidth]{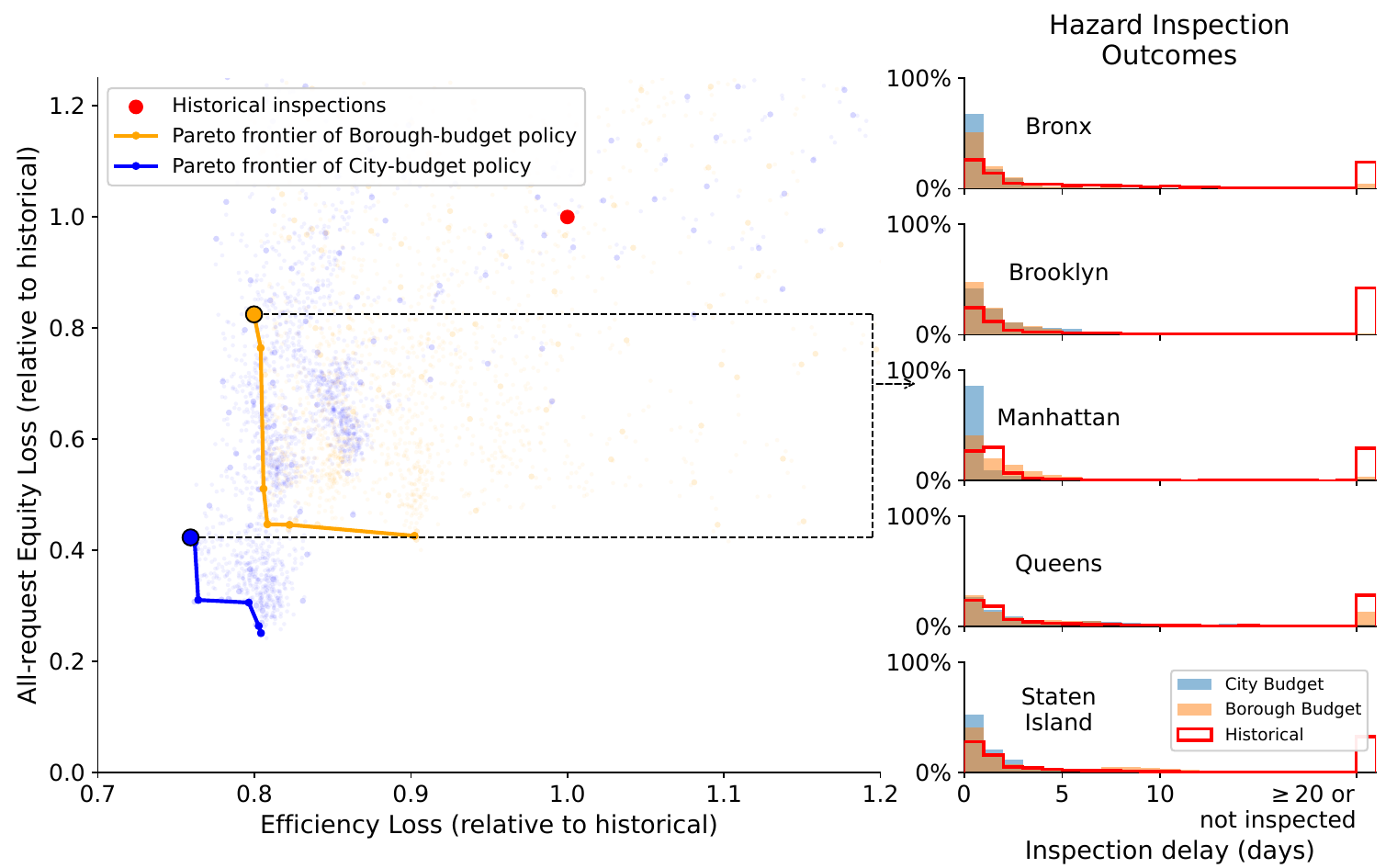}
    \caption{(Left) Pareto frontiers of the equity-efficiency trade-off, under different classes of policies, compared with efficiency and equity loss functions calculated using historical inspections. Lower values on each axis indicate better performance: more efficient policies lie farther left, and more equitable policies lie lower. (Right) Comparing inspection delays for Hazard incidents in different Boroughs, under the most efficient Borough-budget and city-budget policies.}
    \label{fig:pareto}
\end{figure}

\subsection{Results}
\noindent We now analyze optimal policies within each class. \Cref{tab:robustnesstab} contains, for each of 2019, 2021, and 2022, the efficiency and equity loss of three policies compared with historical outcomes: the most efficient and the most equitable Borough budget policies, and a more balanced policy, as calculated using 2019 data. \Cref{tab:budgetalloc} details how these policies allocate budgets across Boroughs (where we include historical outcomes only for 2019). \Cref{fig:costmap} contains the cost borne by each neighborhood (census tract) in New York City in 2019 according to each of these policies. \Cref{fig:pareto} contains the Pareto frontier for each policy class, in terms of efficiency and equity loss. We now overview insights from the analysis. Finally, App. \Cref{tab:borough-sla-outcomes} shows how these policies prioritize different categories and the resulting realized inspection delays and inspection fractions in the five Boroughs in 2019.

\paragraph{Optimal Borough budget policies perform well and are robust.} \Cref{tab:robustnesstab} shows, for the class of Borough budget GPS policies, the most efficient and equitable policies (those with the lowest objective value with $\gamma = 1$ and $\gamma = 0$, respectively). We find that the optimal policies are robust to out-of-sample evaluations: all three policies outperform historical outcomes in both equity and efficiency in future years, and numeric values are similar, though their performance ordering changes. Crucially, there exist (several) balanced policies that outperform historical outcomes in both equity and efficiency.
Further, note that, as shown in \Cref{fig:costmap}, the cost in most census tracts (finer-grained geographical units than Boroughs) can be reduced substantially through such policies, even though the efficiency gain is not substantial at the Borough level, which is partially a consequence of using median inspection delays as the target metric. Comparing optimal budget allocations in \Cref{tab:budgetalloc} with historical inspection shares, we find that the historical budget allocation is closer to the efficient than the equitable endpoint. The most equitable policy, e.g., allocates much more budget to Brooklyn; it historically has longer review periods (\Cref{app:supp_data}).

\paragraph{There is a substantial equity-efficiency trade-off, but equity gains are possible with relatively small efficiency losses.} \Cref{fig:pareto} shows the equity-efficiency Pareto frontiers under different classes of policies. Restricted to the class of Borough budget policies, we see that the price of equity in the application is non-negligible, and the overall trade-off between the two desiderata is large, as expected from \Cref{prop:costofequity}. This may be because of the large differences between Boroughs and the small inspection budget; as noted in \Cref{app:times2budget}, when we re-evaluate the same policies with 1.25 times the historical inspection capacity, the equity-efficiency trade-off becomes substantially smaller in absolute terms, as theoretically predicted. Furthermore, while the full trade-off is substantial, large improvements in equity are possible with relatively small decreases in efficiency. \Cref{tab:robustnesstab} indicates that the most equitable policy is 12.8\% less efficient than the most efficient policy, while achieving a 48.3\% equity gain. We also find many policies that Pareto-dominate historical inspections, and crucially, many of these policies achieve relatively larger equity gains than efficiency gains, pointing to the potential for city agencies to improve the equity of their operations.

\paragraph{Optimal Borough budget policies are close to optimal city budget policies -- the gain from centralization is smaller than the price of equity.} We next compare Borough policies with city ones -- to evaluate an ongoing change that will partially centralize operations. We find that centralizing budgets provides about a 5.1\% efficiency gain (relative to the Borough-efficient policy) and further reduces the equity loss. Though non-negligible, this improvement is modest compared to allocating the budgets and managing the queues within Boroughs more effectively. These findings are in line with the preceding centralization analysis, which showed that the gain from centralization should be small compared to the equity-efficiency tradeoff from how budgets and queues are managed. Similar insights emerge for the equity loss. While the gain from centralization is relatively small compared to optimally chosen Borough-budget policies, we note that centralizing may make these budgets easier to manage and adjust over time, and yield other potential benefits; we discuss these considerations in the appendix. Finally, the comparison of the two may also depend on other model parameters. In \Cref{fig:pfvarsla}, we display the same Pareto frontiers for different model parameters; in one case, the gain from centralization is comparable to the range of the equity-efficiency frontier. 

\paragraph{Gains are robust to different category weights.} As discussed above, our simulation exercise requires choosing a set of priority weights for each category. While these weights were set in consultation with our collaborators at NYC Parks, it may be important to know how dependent our exact results are on them. We conduct this analysis in \Cref{app:priority-weight-sensitivity}.
Broadly, we find that the policies optimized for this set of category priority weights continue to outperform historical outcomes for other sets of weights -- except when weights are almost equal across categories, in which case all policies perform similarly in terms of efficiency. Across weights, we find that our policies, especially the equitable and balanced ones, are substantially more equitable than historical outcomes.

\section{Discussion}
\noindent At a high level, this work frames the design of service level agreements -- a ubiquitous approach to setting and communicating time-sensitive performance guarantees -- as one that connects operational capacity with policy objectives. Compared to standard approaches, it turns the problem from a descriptive one -- what time guarantees have been met historically -- to a prescriptive one, while being aware of capacity constraints. Our approach considers two levers: per-area budgets and incident category prioritization, connecting them to policy objectives based on the service guarantees -- response fractions and delays -- that they induce. 

We develop a theoretical model, formulating a service level agreement design problem. Such a stylized model provides insights on the design space and serves as a foundation for our data-driven simulation-optimization framework. In particular, we analyze the equity-efficiency trade-off and the potential gain from centralizing resources, providing testable predictions for practice. However, this modeling framework omits several aspects important in practice, such as real, non-homogeneous arrival rates and endogenizing the fraction of incidents responded to. Thus, we develop a simulation-optimization framework, to find near-optimal policies computationally, while incorporating such aspects. We apply our simulation framework to the design of SLAs in New York City and find that, in the NYC application, the trade-off between efficiency and equity is indeed not negligible, and we can identify multiple policies on the Pareto frontier that achieve a balance of equity and efficiency objectives and are robust to out-of-sample data. We further observe that by allocating resources well, optimal decentralized agencies perform almost as well as their centralized counterparts, and the gain from centralization is substantially smaller than the price of equity in the application. These simulation findings are in line with our model-based predictions. Our analysis here does not include further challenges important for practice: the \textit{multistage} nature of this system; the choice of the delay fraction guarantee $1-\alpha$; and non-routine service allocation, such as after storms. We discuss these aspects further in the appendix.

\paragraph{Interaction with NYC Parks.} This paper benefited greatly from interaction with NYC Parks, with whom we have an extended collaboration across projects. Their questions about how to design SLAs led to the initial formulation of the work; their input regarding the available levers further contributed to the establishment of the theoretical framework; and their feedback helped us construct category weights that went beyond considering average risk ratings. In \Cref{sec:dprslides}, we further discuss this collaboration and include snapshots of materials we used when communicating with NYC Parks, including an analysis of the historical SLAs met by the agency and the high-level model formulation. These materials highlight the strategy that we find most useful in our communications: first, focusing on application results and summarizing key assumptions, as opposed to details of the theoretical model, better leverages practitioners' insights; second, showing the outcomes of policies and confirming their practicality is a good way to verify the design of objective functions---while the mathematical formulations of the equity and efficiency objectives matter in the optimization, their importance is best illustrated to practitioners by the budgets and allocations they induce (cf. \citet{delarue2024algorithmic}). We hope that our work aids further research in the design of government service allocation policies, in collaboration with practitioners.

\bibliographystyle{plainnat}
\bibliography{ref} 

\newpage
\appendix
\crefalias{section}{appendix}
\crefalias{subsection}{appendix}
\crefalias{subsubsection}{appendix}
\section{Supplementary materials}

AI disclosure: We use LLM coding agents to bug check our code and develop and run new analyses. We also used such tools as proof assistants and checkers, as well as to provide feedback on the writing.

\subsection{Additional modeling  and results discussion}
Our modeling framework may be extended or applied in several ways. 

\paragraph{Design of $\alpha$, the tail-probability violation tolerance for inspected incidents.} In our model, $1-\alpha_{k,b}$ governs the fraction of \textit{responded-to} incidents for which the delay guarantee is met, e.g., 90\% of non-urgent incidents, that are inspected, will be inspected within some number of days. In this work, we do not consider optimizing this number -- we do not treat it as a decision variable in either the theoretical section or in the data-driven simulation, and we view it as a constant across types. In our primary simulations, we set $1-\alpha = .5$ and conduct robustness analyses with $1-\alpha = .75$ in the appendix. To our knowledge, homogeneous, non-optimized $\alpha$ reflects SLA practice: while the delay $z_{k,b}$ is highly heterogeneous across types, $\alpha$ is not; see, e.g., the policy documents for each city linked in the introduction: in each example, $z$ may differ across types, but $\alpha$ does not (note that the policy documents state numbers in terms of overall incidents, not responded to incidents, which in our model can be found by multiplying $1-\alpha_{k,b}$ and the inspected fraction). However, additionally optimizing for $\alpha$ may be an interesting question for future work, and it would be valuable to systematically understand how it changes achievable SLAs. 

\paragraph{Multistage resource allocation.} Incident response is often a multistage pipeline: after a reported incident, a trained inspector may travel to the site to understand its severity and urgency; given this data, a maintenance crew may be scheduled and ultimately dispatched to resolve the issue. Agencies must first decide whether the stated SLAs will be for the overall end-to-end response (report to incident being physically addressed), each stage separately, or just the first stage. Following current agency practice, in this work we tackle the simplest version of this question by simulating first-stage (inspection) allocation. However, according to our agency contacts, this choice itself is up for discussion, and it would be interesting in future work to model the full end-to-end complexity of the system. 

How to best model and optimize such multistage decisions is an interesting methodological question. Note that resources (e.g., inspectors vs maintenance crews) differ across stages. Furthermore, incident \textit{types} also differ. In the first stage, before an inspection, the agency often only has coarse information on each incident (e.g., category); it does not yet have access to incident risk. After the inspection, however, it has detailed information that is used to schedule maintenance response: the exact risk levels and severity as observed by the trained inspector. Effective modeling of multistage decisions and SLAs thus requires consideration of both downstream resources and how incident types may distributionally change (now, we are not just concerned with an incident's mean priority level, but also the probability that it is a more severe incident requiring urgent maintenance). 

\paragraph{Non-routine (emergency) periods.} Our model and simulation analyses focus on ``non-emergency'' periods, {using historical data to represent fluctuations in incident arrivals and inspections}. As shown in App. \Cref{fig:srs_ins}, during storm periods, such as in August 2020, service requests and inspections may increase tenfold compared to the same time in other years. The ability to capture such ``emergency'' periods, where a surge of incidents arrives and auxiliary measures need to be taken into account when scheduling inspections, is a valuable extension of practical significance. We note that the existence of such storm periods may increase the bureaucratic value of centralization: even if resources are not cross dispatched across areas in routine periods, having centralized control and operations of all resources in a city may help in logistics during such emergency periods, when resources are cross-dispatched. 

\paragraph{The cost of dropped incidents.} From \Cref{tab:borough-sla-outcomes}, our optimized solutions -- alongside historical choices in many cases -- drop most incidents (up to 80\%--90\%) of some types (e.g., \textit{Prune}) in many Boroughs, and conversely inspect most (almost $90$+\%) incidents of other types, such as \textit{Hazard}. In contrast, historical decisions are more balanced -- about 80\% of Hazards are inspected, alongside much more of the other types (though this varies across Boroughs substantially). 

Our analysis shows that this tradeoff is perhaps inevitable, and our solutions are the outcome of optimizing (using Bayesian optimization search) the given objective functions. If a policymaker deems this solution undesirable, then one response would be to change the objective function. However, this cannot come for free, with corresponding increases in delays for other incident types. (We further note that it may not be trivial to do so; e.g., in simulations where we increased the cost $D$ of dropping an incident, it approximately made drop rates even more extreme, with lower weight incidents being dropped even more. The second part of the Cost term, for dropped incidents, likely needs to be nonlinear with higher dropped fractions having greater marginal cost; one could also add additional constraints directly). Which solution is preferable is up to policymakers. We view such a finding as a use case of an analysis such as ours: as prior work has shown, choosing objective functions is challenging, and often requires seeing solutions to realize what is missing from the stated functions. 
In practice, in our conversations with the agency, they indicated that they navigate this issue with special focus periods with additional resources. For example, they may have several months in which their goal is to close out many such pending low priority incidents, with additional capacity. We view such decisions as outside of our modeling framework.

\subsubsection{Supplementary information on collaboration with NYC Parks.}
\label{sec:dprslides}

This project was directly inspired by conversations with NYC Parks. The research team was collaborating with them on other projects \citep{liu2023quantifying}, including co-publishing with them on understanding crowdsourced reporting behavior: do some neighborhoods report problems at slower rates than others; in short, we found that the answer was yes, correlating with neighborhood demographics. That project led to further conversations, and eventually this project and several others, including one in which we developed an integer program to help optimize their upcoming 9-year tree planting cycle schedule (every block is visited in a 9-year cycle to determine whether trees should be planted there, and the order of visiting blocks has substantial needs-based desiderata and geographic balance objectives).

The team at NYC Parks was (and still is) in the process of (1) centralizing bureaucratic control and visibility of resources, which would enable but not require more effective cross-dispatch across locations; and (2) overhauling the SLAs that are published online, given that they currently do not reflect their practice (as discussed in the introduction). Conversations regarding the challenges they face in designing these SLAs led us to formulate this research project, especially given further evidence that this is a general question faced by many agencies across cities. To date, they have not finished their SLA design process, and our insights have not been incorporated. However, we have presented to the team, including high-level insights on the equity effects of historical allocations compared to calculated ones. 

\begin{figure}[htb]
    \centering
    
    \begin{subfigure}[b]{0.32\textwidth}
        \centering
        \includegraphics[width=\textwidth]{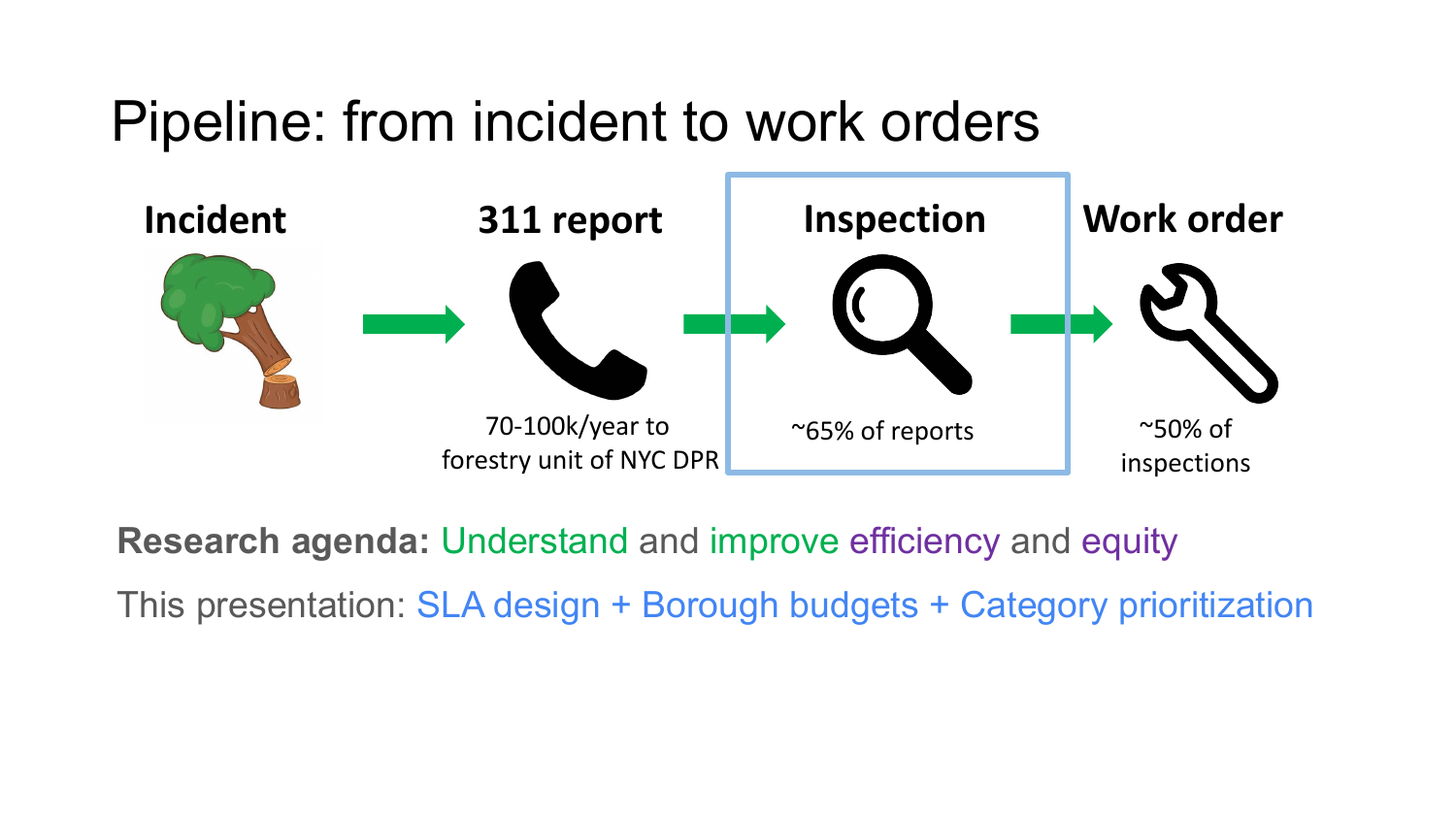}
        \caption{Research questions}
        \label{fig:slide1}
    \end{subfigure}
    \begin{subfigure}[b]{0.32\textwidth}
        \centering
        \includegraphics[width=\textwidth]{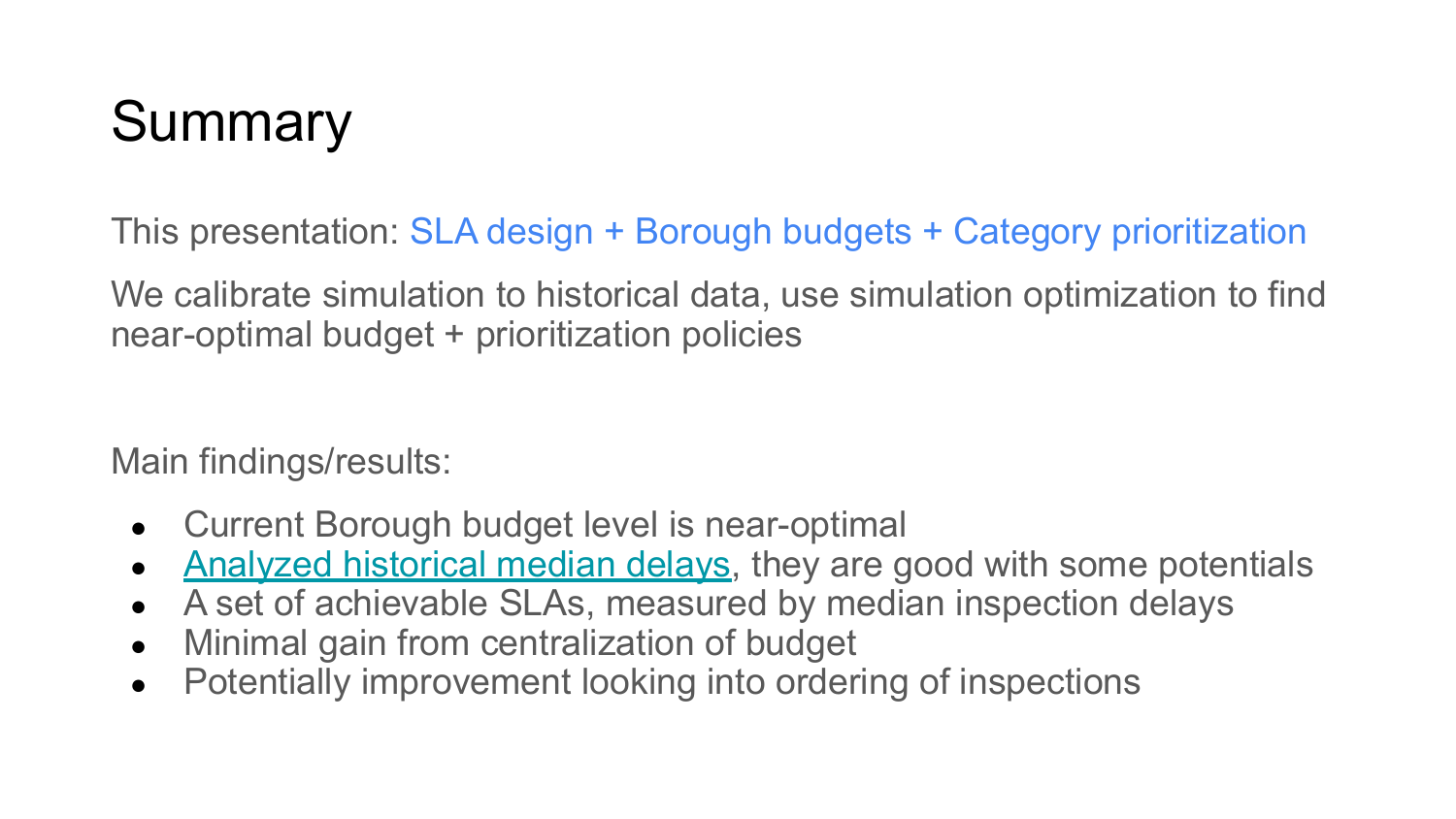}
        \caption{Summary of findings}
        \label{fig:slide2}
    \end{subfigure}
    \begin{subfigure}[b]{0.32\textwidth}
        \centering
        \includegraphics[width=\textwidth]{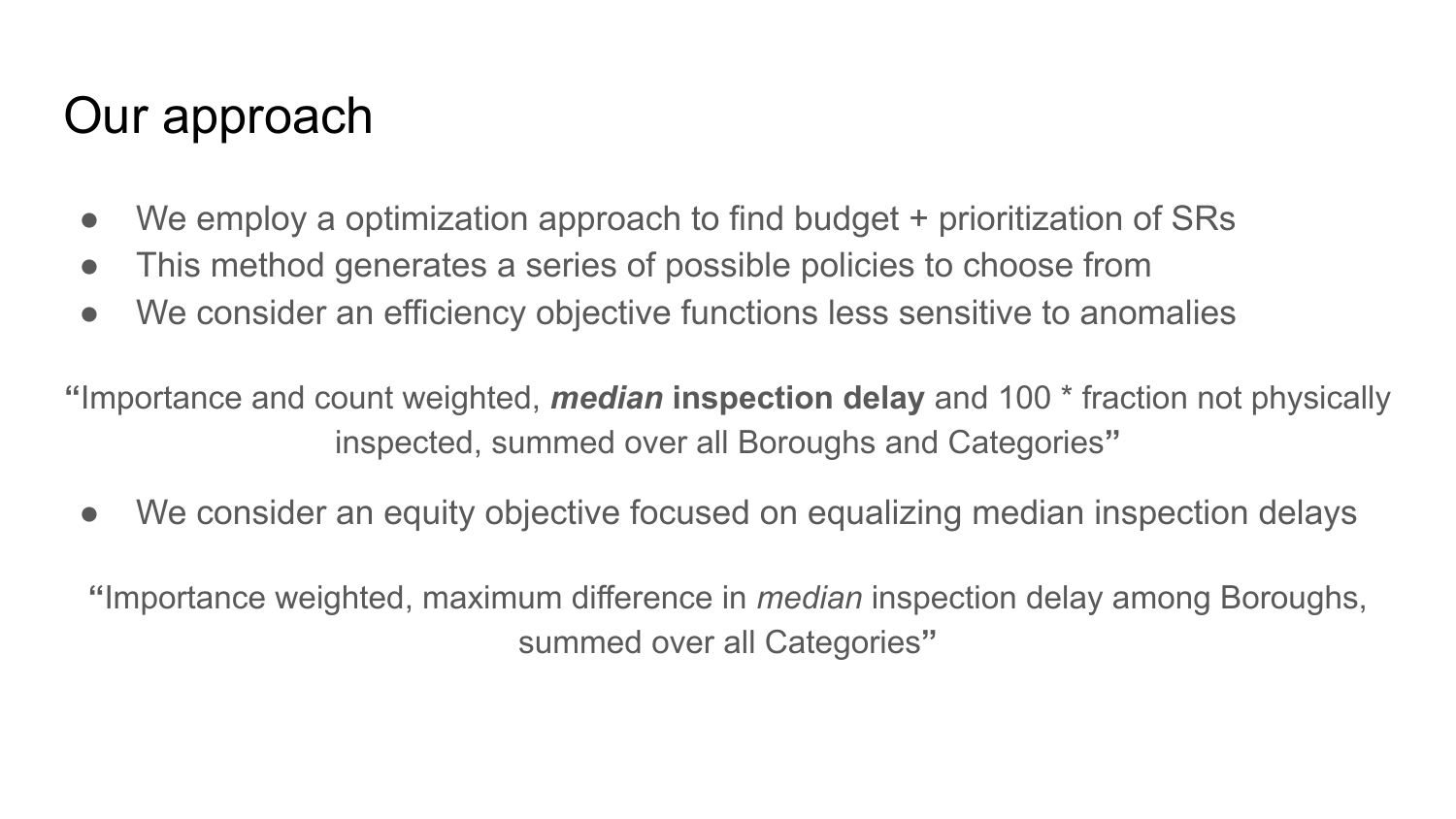}
        \caption{Summary of methods}
        \label{fig:slide3}
    \end{subfigure}
    \hfill
    \begin{subfigure}[b]{0.32\textwidth}
        \centering
        \includegraphics[width=\textwidth]{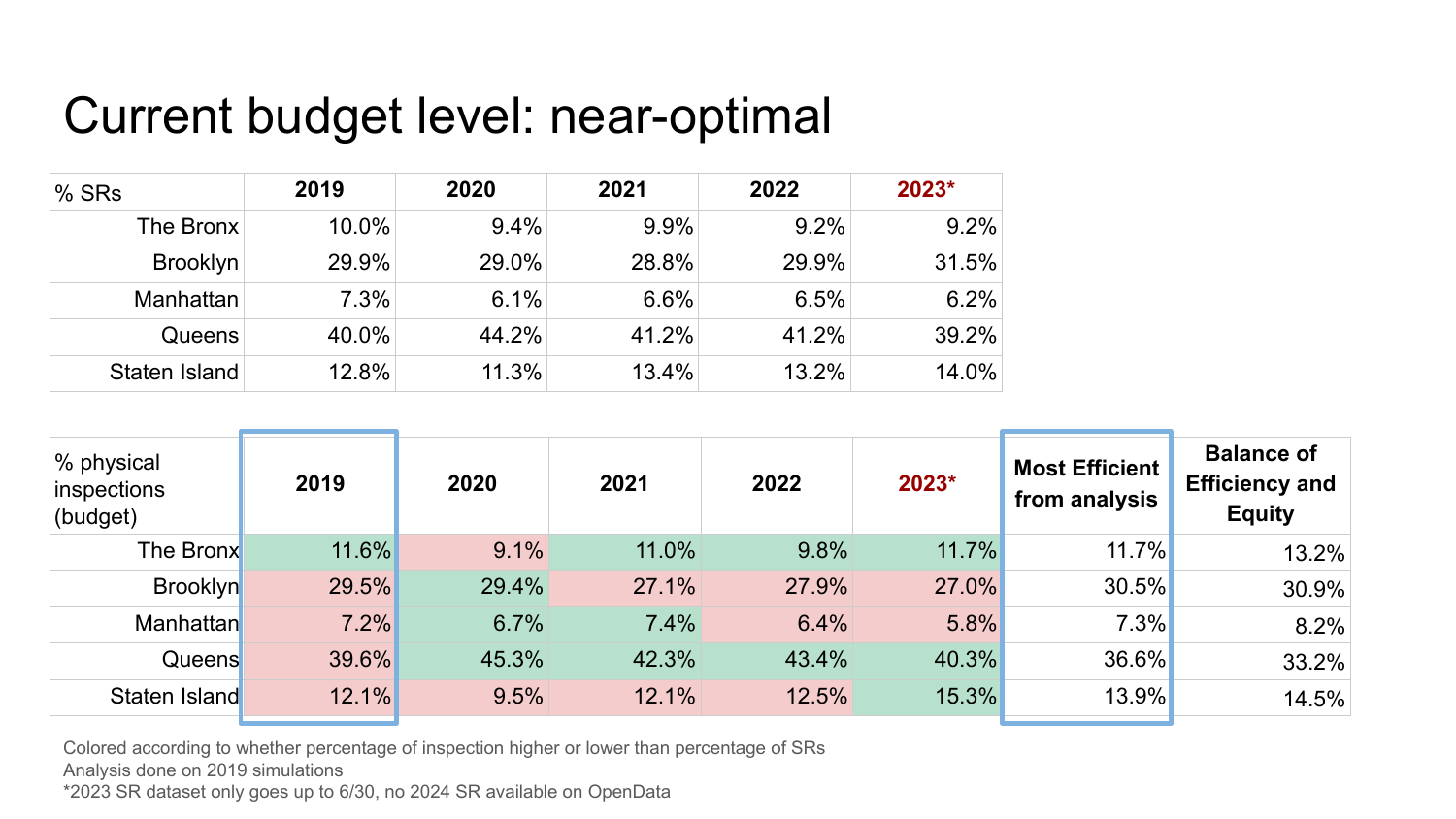}
        \caption{Results on budget allocation}
        \label{fig:slide4}
    \end{subfigure}
    \begin{subfigure}[b]{0.32\textwidth}
        \centering
        \includegraphics[width=\textwidth]{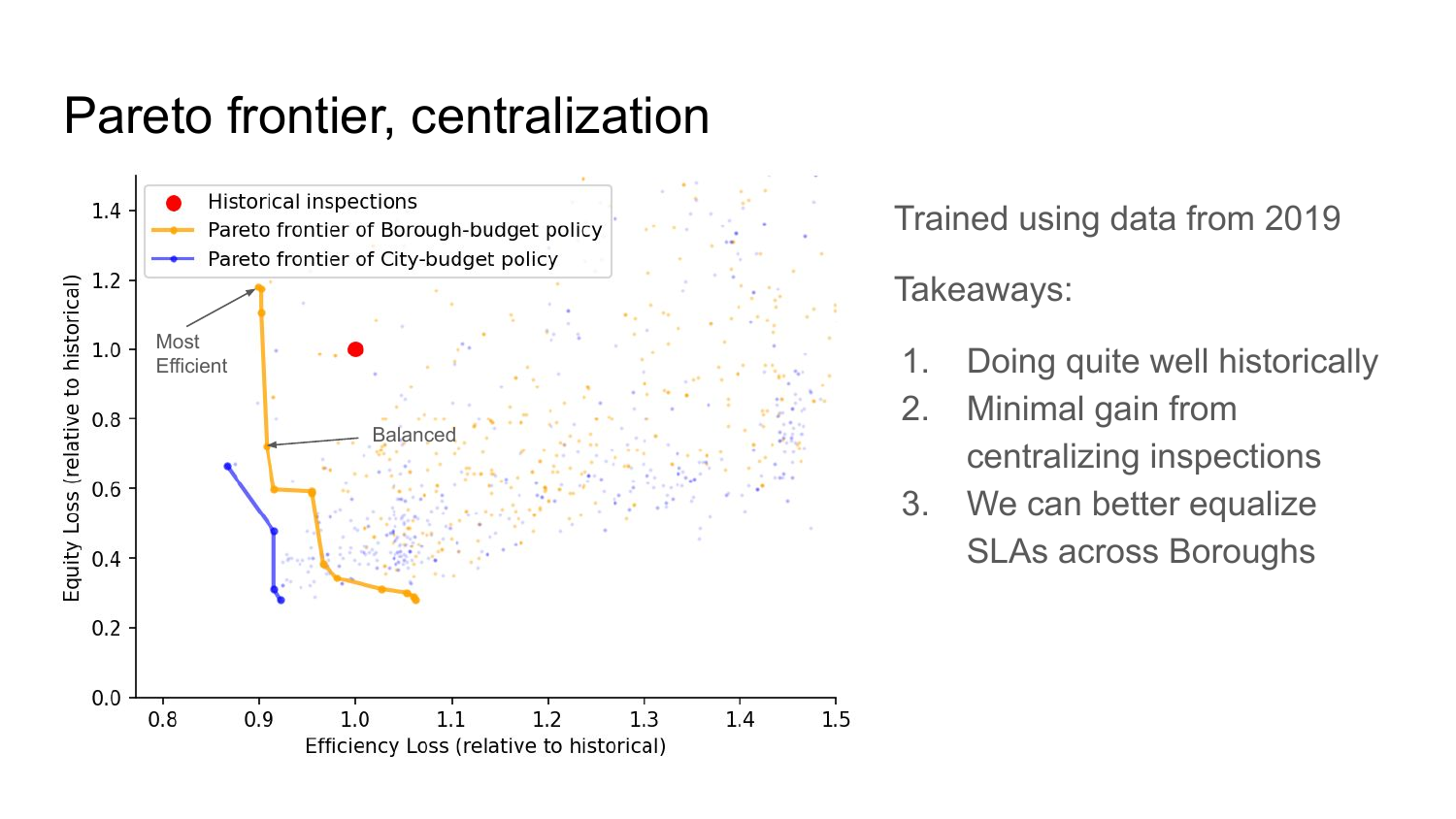}
        \caption{Results on Pareto frontier}
        \label{fig:slide5}
    \end{subfigure}
    \begin{subfigure}[b]{0.32\textwidth}
        \centering
        \includegraphics[width=\textwidth]{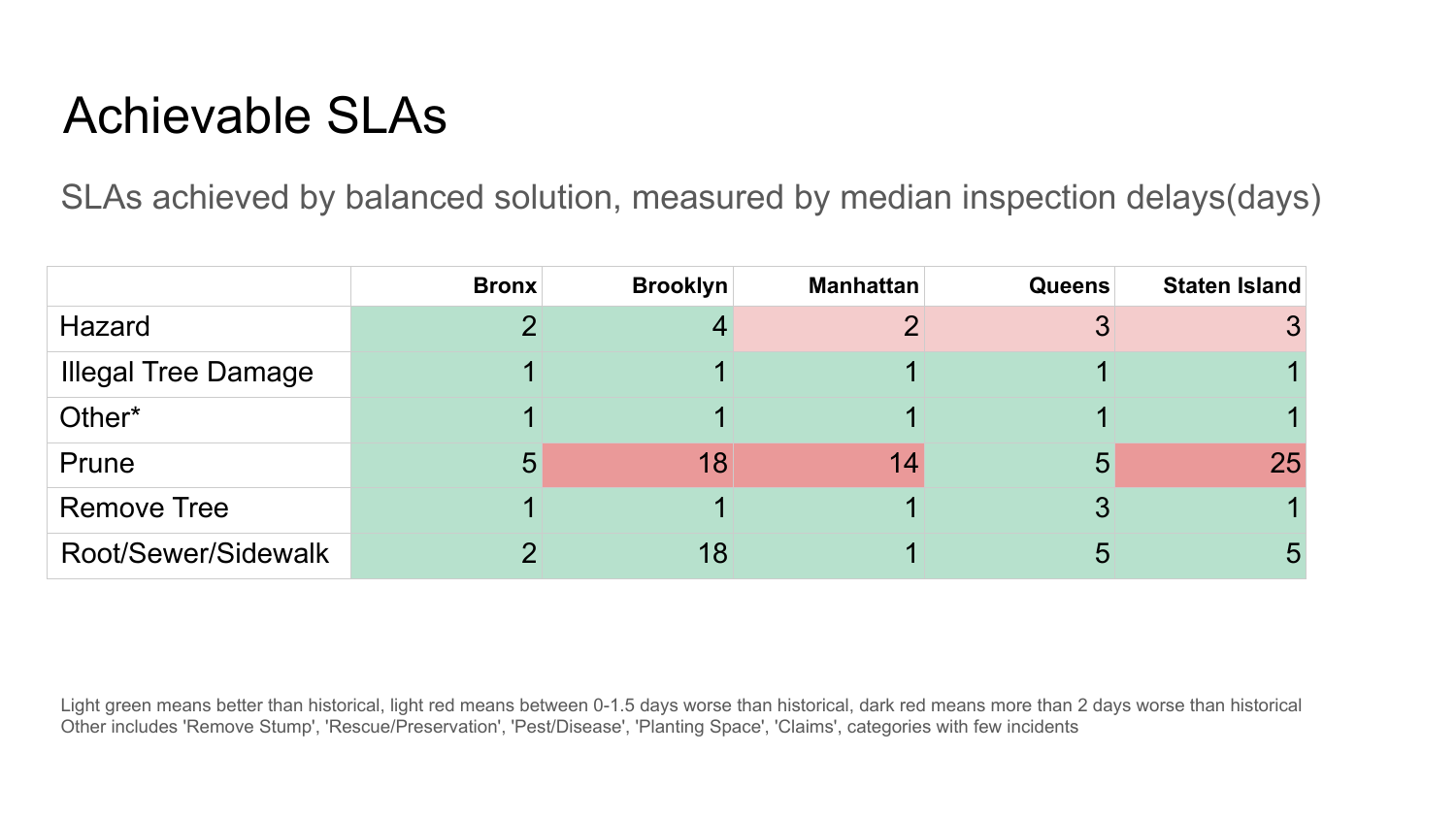}
        \caption{Results on achievable SLAs}
        \label{fig:slide6}
    \end{subfigure}
    \hfill
    
    \caption{Samples of slides used in communicating with NYC Parks on September 27th, 2024.}
    \label{fig:slides}
\end{figure}

\begin{figure}[htb]
    \centering
    
    \begin{subfigure}[b]{0.48\textwidth}
        \centering
        \includegraphics[width=\textwidth]{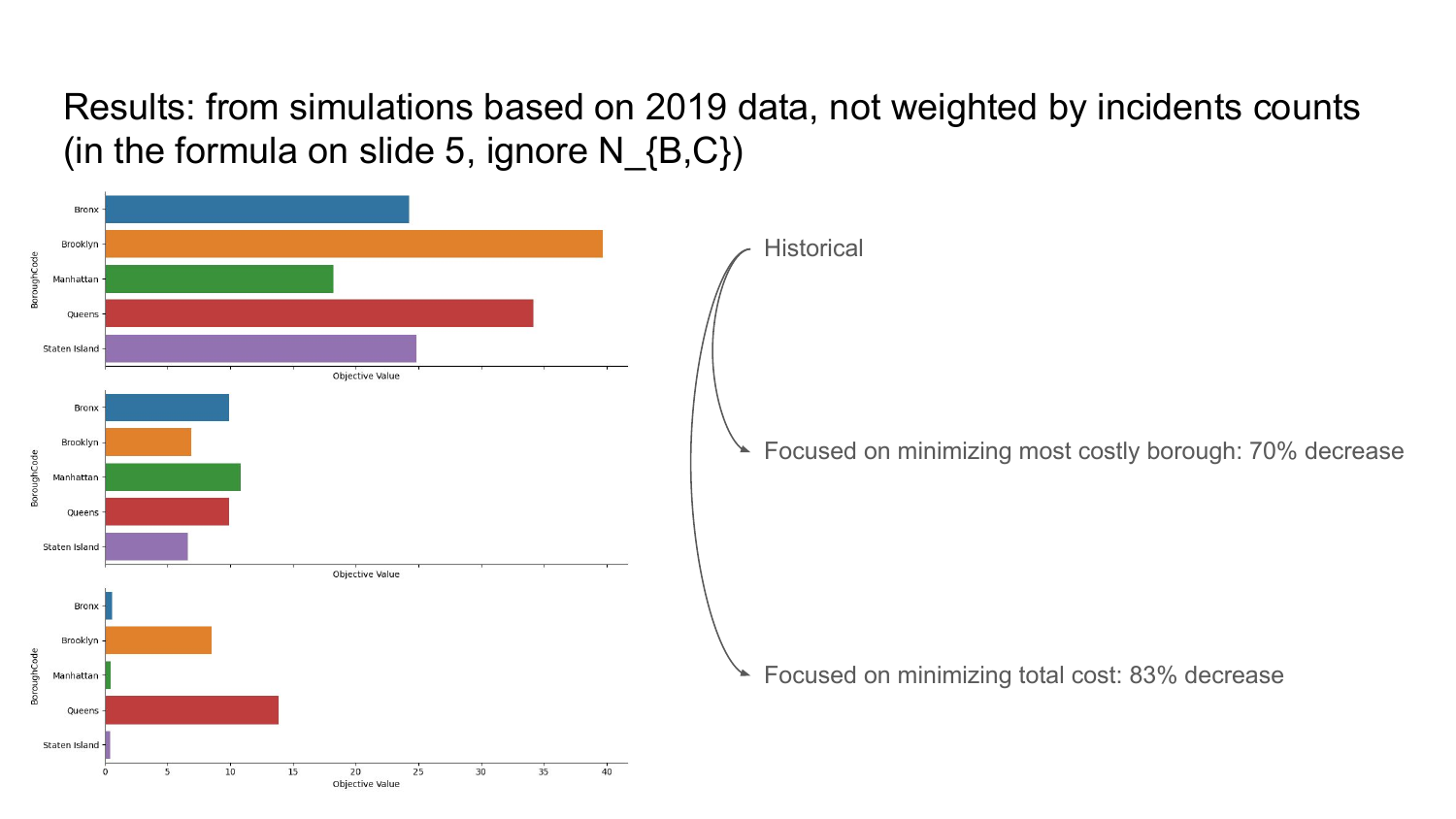}
        \caption{}
        \label{fig:slide1old}
    \end{subfigure}
    \hfill
    \begin{subfigure}[b]{0.48\textwidth}
        \centering
        \includegraphics[width=\textwidth]{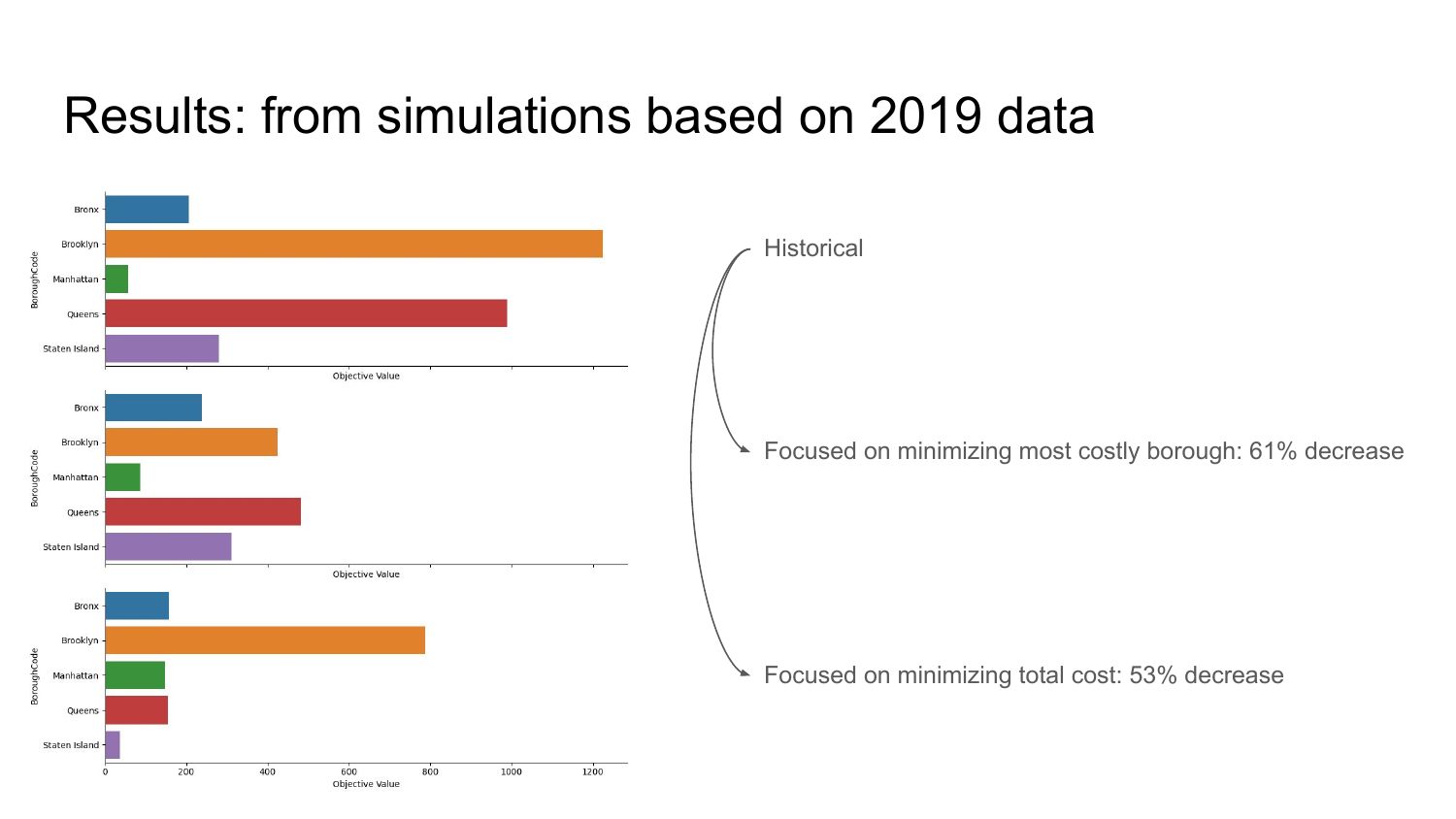}
        \caption{}
        \label{fig:slide2old}
    \end{subfigure}    
    \caption{Samples of slides used in communicating with NYC Parks on March 28th, 2024, showing different effects of optimal policies under different objective functions.}
    \label{fig:slidesold}
\end{figure}

\begin{figure}
    \centering
    \includegraphics[width=1.0\linewidth]{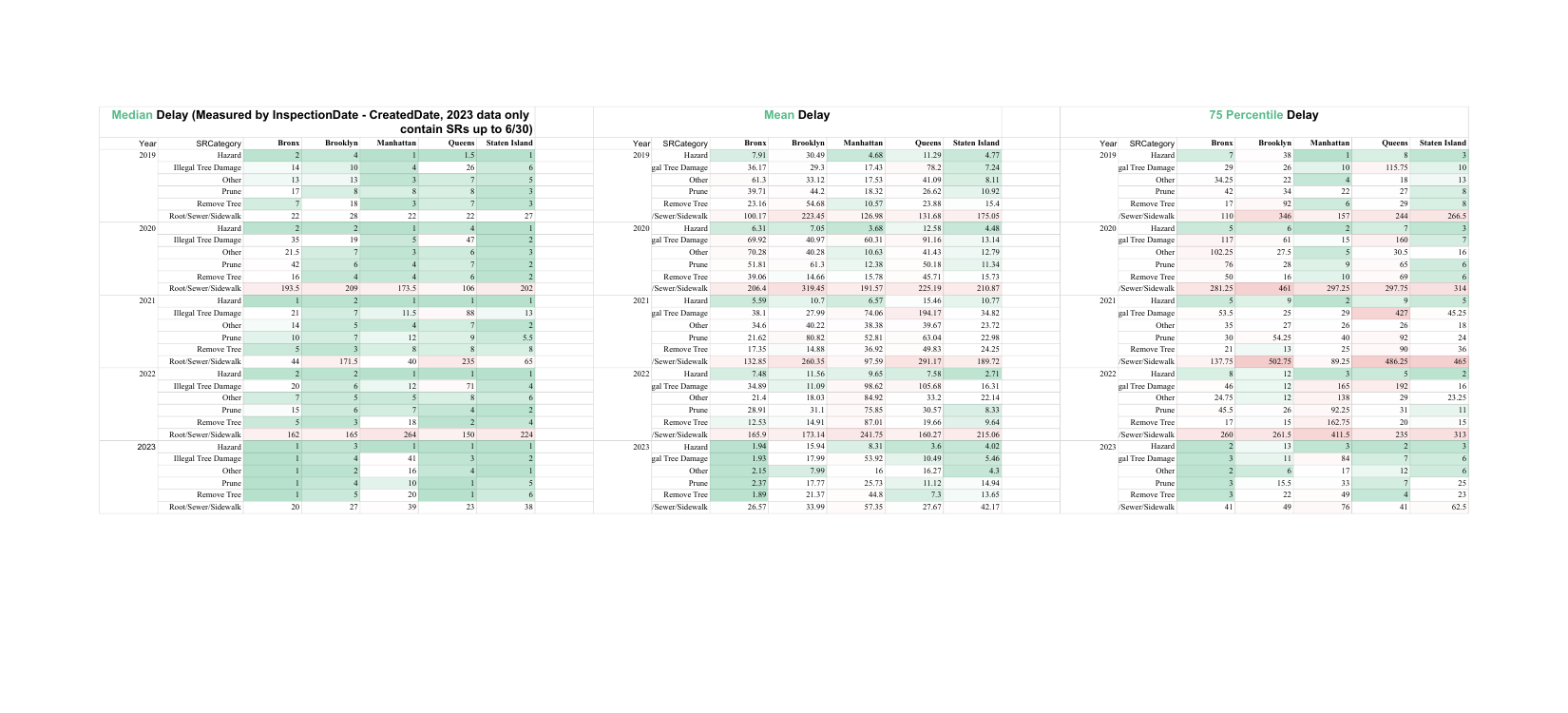}
    \caption{Snapshot of a spreadsheet we shared with NYC Parks and used to audit their historical SLAs under different definitions, grouped by Borough, category, and time period.}
    \label{fig:slaspreadsheet}
\end{figure}

Below, to give context for other academic-practitioner collaborations, we provide snapshots of materials used in our rounds of communication with NYC Parks. \Cref{fig:slides} shows a sample of the slides we used when presenting to NYC Parks in September 2024; \Cref{fig:slidesold} shows a sample of the slides we used when presenting to NYC Parks in March 2024; \Cref{fig:slaspreadsheet} shows a sample spreadsheet that we shared with NYC Parks on auditing their historical SLAs. These slides showcase our effort in using feedback from NYC Parks practitioners on the outcomes of optimal policies to identify ``correct'' objective functions to use. We believe that such back-and-forth -- of presenting potential solutions and eliciting objective functions -- is a central component of such collaborations, where domain experts may intuitively understand what makes for a good solution, but may not be able to translate it to specific objective functions absent seeing their implications in terms of optimization outputs. 

Finally, we note that several structural features of our context aid such a collaboration. Most importantly, in this project, we do not require the use of any private data -- all data used is publicly available via the NYC Open Data Portal. This NYC culture (backed by city law) of making agency data available in near real time has several beneficial implications: first, data access is not a bottleneck, and we can publicly share data with substantially lower data-access and privacy barriers. Second, that this data is already publicly available means that agencies have a corresponding technical infrastructure and expertise around such data, contributing to organizational sophistication around data to inform operations.  Thus, collaboration with agency data scientists can focus on communication of domain expertise and understanding the data, as opposed to legal and technical hurdles in sharing it.

That being said, translating our analysis to impact is a separate challenge. On one level, based on practitioner feedback, it has already been a useful contribution of our framework to trace out a Pareto frontier, and show how resource constraints matter across categories and areas. Similarly, they have communicated appreciation of the descriptive analyses of historical inspection decisions and SLA adherence, with comparisons to more equitable allocations. On another level, actual \textit{adoption} of SLAs based on this work requires substantial further steps -- obeying regulatory and other policy constraints, modeling other components such as limited (and potentially) costly cross-Borough dispatch, and further organizational buy-in. 

\subsection{Supplementary data-driven simulation approach and results information}
\label{app:supp_data}
This section provides further detail on simulation inputs, simulation process, supplementary results, and robustness analyses that support the main data-driven analysis.

\subsubsection{Data and other inputs.} 
\begin{figure}[htb]
    \centering
    
    \begin{subfigure}[b]{0.48\textwidth}
        \centering
        \includegraphics[width=\textwidth]{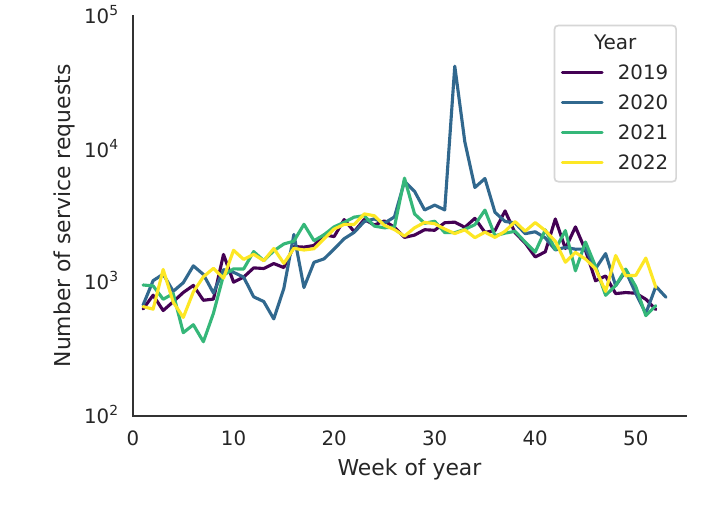}
        \caption{{Service (inspection) requests}}
        \label{fig:srs}
    \end{subfigure}
    \hfill
    \begin{subfigure}[b]{0.48\textwidth}
        \centering
        \includegraphics[width=\textwidth]{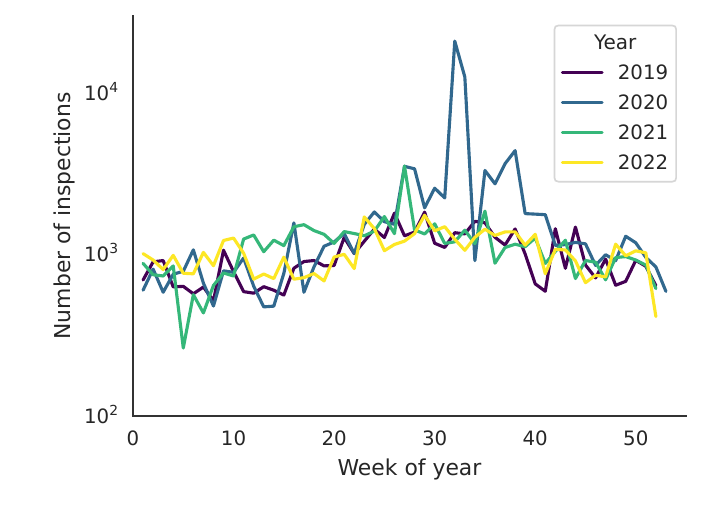}
        \caption{{Inspections}}
        \label{fig:ins}
    \end{subfigure}
    \caption{{Number of service (inspection) requests and inspections by week from 2019 to 2022.}}
    \label{fig:srs_ins}
\end{figure}

\begin{figure}[htb]
    \centering
    \includegraphics[width = .6\textwidth]{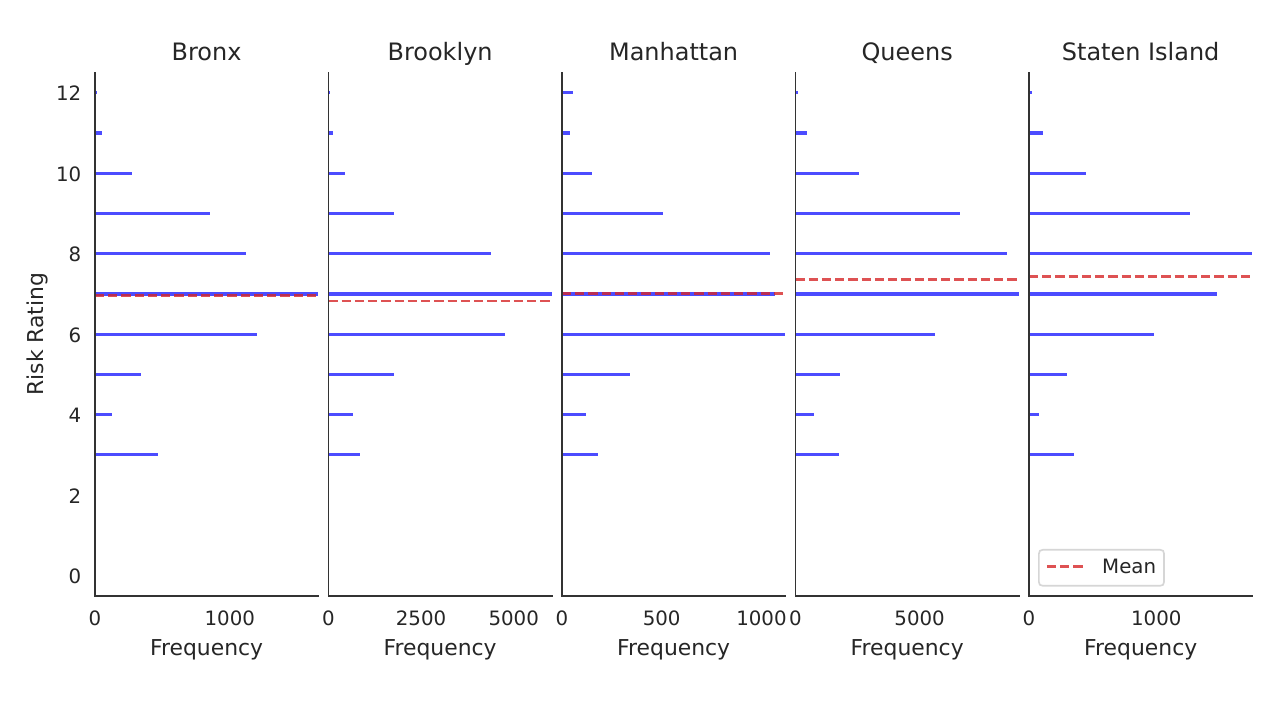}
    \caption{Distribution of risk ratings for Hazard incidents across Boroughs, with means of risk ratings within each Borough marked.}
    \label{fig:riskdist_hazard}
\end{figure}

The data used are postprocessed from publicly available service requests (\url{https://data.cityofnewyork.us/Environment/Forestry-Service-Requests/mu46-p9is/}) and inspections performed (\url{https://data.cityofnewyork.us/Environment/Forestry-Inspections/4pt5-3vv4/}). \Cref{fig:srs_ins} shows {the number of service requests received and the number of inspections performed each week across all categories}, from 2019 to 2022. Large seasonality and variance render assumptions of Poisson arrivals unjustified. {\Cref{tab:number_srs_by_borough_year,tab:number_inspections_by_borough_year} report {annual counts} by Borough used in our simulations.} \Cref{tab:avgrisk} provides average risk ratings of each category of incidents in each Borough, and \Cref{fig:riskdist_hazard} shows the distribution of risk ratings of Hazard incidents across Boroughs. We find that (1) the difference in risk ratings among different categories is insufficient to reveal the practical difference in priorities that NYC Parks would like to assign to these categories, and (2) the difference in aggregate risk ratings of the same category of incidents among different Boroughs is too small to justify treating Boroughs differently. Thus, in our main application, we use risk levels that better reflect the importance of each category, uniform across all Boroughs.

Hyperparameter values introduced in \Cref{sec:sim_given_policy} are as follows. The hyperparameter $\rho$, controlling deviations from FCFS inspection, is calibrated to be 0.15. We note that this means the simulations are much closer to strict FCFS ($\rho=0$) than to random inspection ($\rho=1$). We note that FCFS is intended: currently, each inspector faces a dashboard that sorts incidents by category and then by age when making inspection decisions. However, of the \num{102560} Hazard incidents that were both reported and inspected from 2019 to 2023, \num{29967}, or 29.2\% of them were inspected later than at least one same-Borough Hazard incident that was reported later, indicating that FCFS is indeed not perfectly implemented in practice. The hyperparameters $\tau_b$, controlling review frequency for each Borough, are calibrated to be the following values: the Bronx 28 days, Brooklyn 80 days, Manhattan 26 days, Queens 27 days, Staten Island 36 days. This hyperparameter is useful for calibrating the simulator to endogenous disparities among Boroughs, for example Brooklyn having longer tails in historical inspection delays, as shown in the Brooklyn rows of App. \Cref{tab:borough-sla-outcomes}.

\begin{table}[tb]
\small
\centering
\begin{tabular}{lrrrrr}
\toprule
 & Bronx & Brooklyn & Manhattan & Queens & Staten Island \\
\midrule
2019 & 7512 & 22433 & 5514 & 30004 & 9613 \\
2020 & 13024 & 40385 & 8420 & 61471 & 15784 \\
2021 & 8161 & 23759 & 5415 & 33967 & 11079 \\
2022 & 7044 & 22969 & 5004 & 31627 & 10156 \\
\bottomrule
\end{tabular}
\caption{{Number of requests in each Borough in each year (for the categories in our simulation). Requests linked to multiple inspections are duplicated.}}
\label{tab:number_srs_by_borough_year}
\end{table}

\begin{table}[tb]
\small
\centering
\begin{tabular}{lrrrrr}
\toprule
 & Bronx & Brooklyn & Manhattan & Queens & Staten Island \\
\midrule
2019 & 5636 & 13304 & 3907 & 19509 & 5984 \\
2020 & 9243 & 31979 & 6654 & 44951 & 9390 \\
2021 & 6736 & 14462 & 3993 & 23459 & 7571 \\
2022 & 4611 & 14451 & 3119 & 24100 & 5695 \\
\bottomrule
\end{tabular}
\caption{{Number of completed inspections by Borough in each year (for the categories in our simulation).}}
\label{tab:number_inspections_by_borough_year}
\end{table}

\begin{table}[tb]
\small
\centering
\begin{tabular}{lrrrrrr}
\toprule
 & Bronx & Brooklyn & Manhattan & Queens & Staten Island & Assigned category priority weights\\
\midrule
Hazard              & 6.96 & 6.82 & 7.01 & 7.36 & 7.44 & 10\\
Illegal Tree Damage & 5.91 & 5.18 & 5.46 & 6.69 & 6.47 & 8 \\
Other               & 6.86 & 6.19 & 6.18 & 7.67 & 7.75 & 6\\
Prune               & 5.42 & 6.62 & 6.41 & 7.16 & 6.67 & 4 \\
Remove Tree         & 6.69 & 6.73 & 5.85 & 7.21 & 7.14 & 8\\
Root/Sewer/Sidewalk & 5.26 & 4.54 & 5.14 & 4.24 & 4.45 & 4\\
\bottomrule
\end{tabular}
\caption{Mean observed risk rating among linked inspected-and-rated rows by incident category and Borough, 2019--2022, alongside the assigned category priority weights used in the loss functions. Recorded risk ratings and assigned policy weights are distinct objects; the assigned weights are model inputs, not request-level ex ante measurements.}
\label{tab:avgrisk}
\end{table}

\FloatBarrier
\subsubsection{Supplementary information on simulation process.}

\label{sec:alt_policy}

In this subsection, we provide more details about the simulation process used within our simulation analyses. To decide the counterfactual number of inspections in each Borough on day $t$, we distribute {the historical citywide inspection count for the relevant categories} $I^t$ among Boroughs according to a Multinomial distribution governed by the budget allocation $\{C_b\}$:
\[
    \{I_b^t\}\sim\operatorname{Multinomial}(I^t,\{C_b\}),
\]
where $I_b^t$ denotes the number of inspections allocated to Borough $b$ on day $t$. Thus, over many days, Borough $b$ is allocated approximately the fraction $C_b$ of offered inspection capacity while allowing for day-to-day fluctuations.

Once $I_b^t$ is determined, we allocate inspections across the categories with nonempty backlogs in that Borough. If $I_b^t$ is at least the Borough's total backlog, every pending request in that Borough is inspected and any remaining Borough capacity is unused rather than transferred to another Borough. Otherwise, let $\mathcal K_b(t)$ denote the set of categories with nonempty backlogs and let $\Phi_b(t)=\sum_{k\in\mathcal K_b(t)}\phi_{k,b}$. The initial category allocations are drawn according to
\[
    \{I_{k,b}^t\}\sim
    \operatorname{Multinomial}\left(I_b^t,
    \left\{\frac{\phi_{k,b}}{\Phi_b(t)}\right\}_{k\in\mathcal K_b(t)}\right),
\]
the discrete multinomial analogue of allocating Borough capacity in proportion to the active GPS weights. Allocations exceeding a category's backlog are capped, and excess inspection slots are reallocated among categories with remaining backlog in proportion to the same weights.

The following pseudocode provides an overview of the simulation process for Borough-budget policies.

\SetAlgorithmName{Simulation process}{simulation}{}
\RestyleAlgo{ruled} 
\begin{algorithm}[h]
\footnotesize
\caption{\footnotesize Simulation adapted to historical data under the Borough-budget policy framework.}
 \KwData{number of days $T$; {historical arrivals} $\{\mathcal{N}_{k,b}^{t}\}$; {historical citywide inspections performed on each calendar day} $\{I^t\}$.}
 \KwVars{Borough budgets $\{C_b\}$; GPS scheduling policy $\{\phi_{k,b}\}$;  per-review retention probability $\{p_{k,b}\}$.}
 \KwParams{review period length $\tau_b$; FCFS violation $\rho$.}
 \KwResult{inspected incidents with inspection delay, incidents still in backlog, and incidents dropped.}
 $\text{t} \gets 1,$
 $\text{backlog} \gets \emptyset,$
 $\text{inspected} \gets  \emptyset,$
 $\text{dropped} \gets  \emptyset.$\\
 \While{t $\le T$ }{
 backlog $\gets $ backlog $\cup$ $\mathcal{N}_{k,b}^t${ for all $k,b$}\\
 inspections in each Borough $\{I_{b}^t\} \gets \text{Multinomial}(I^t, \{C_b\})$\\
  
  \For{\text{Borough} $b \in \mathcal{B}$}{
   number of inspections for each category $\{I_{k,b}^t\}$ $\gets$ $\text{Multinomial}(I_{b}^t, \{{\phi_{k,b}}/\Phi_b(t)\})$, capped by category backlog with excess slots reallocated among categories with remaining backlog;\\
   inspections for each category $\gets $ {Borough $b$ }backlog with indices Unif$\left(I_{k,b}^t, \rho(B_{k,b}^t- I_{k,b}^t) + I_{k,b}^t\right)$;\\
   inspected $\gets $ {inspected $\cup$ }inspections for each category;\\
   backlog $\gets$ backlog $\setminus$ inspections for each category;\\
  \If{$t\bmod \tau_b$ = 0 {for Borough $b$}}{
    \For{incident $\in$ {Borough $b$'s }\text{backlog}}{
    with probability $(1-p_{k,b})$, backlog $\gets$ backlog $\setminus$ incident, dropped $\gets$ dropped $\cup$ incident
    }
  }
  }
  {$t\gets t+1$};
 }
\end{algorithm}

Under the city-budget policy, the step in which inspections in each Borough are determined ($\{I_{b}^t\} \gets \text{Multinomial}(I^t, \{C_b\})$) is skipped, and inspections are directly allocated to each Borough--category pair: we maintain a different queue for incidents of each Borough--category combination.

\FloatBarrier

\subsubsection{Supplementary information on simulation results in \Cref{sec:results}.}
\label{sec:appres}
\begin{figure}[htb]
    \centering
    \includegraphics[width=.95\textwidth]{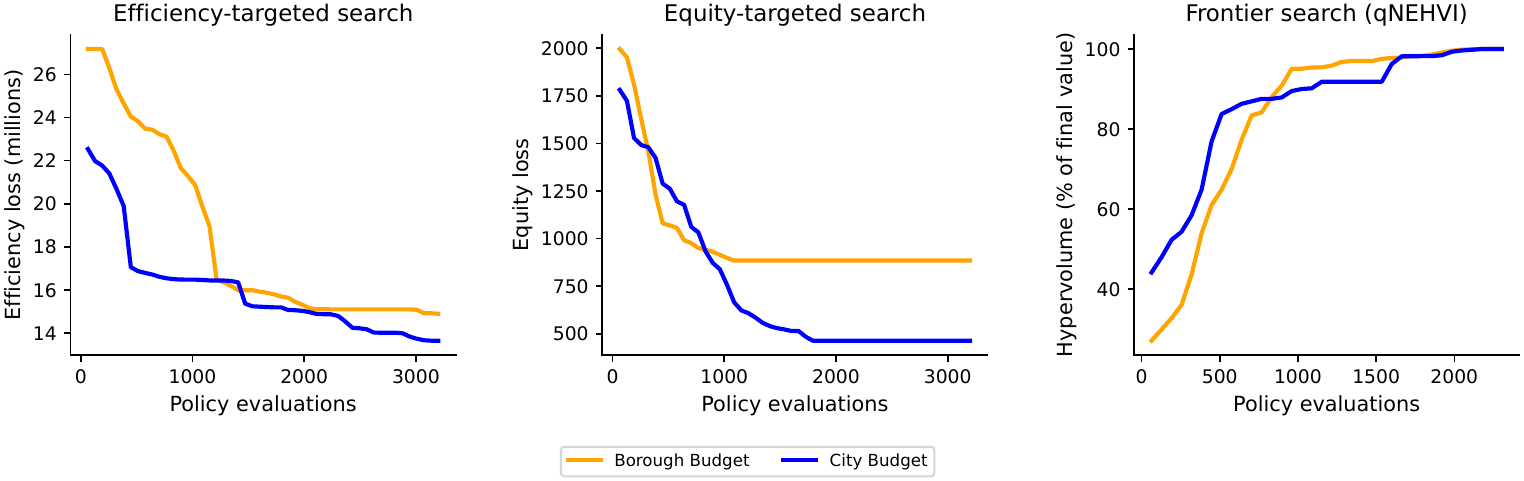}
    
    \caption{Search progress for the primary Borough- and City-budget Bayesian optimizations. The first two figures report the best loss observed in the efficiency- and equity-targeted searches (lower is better); the third reports the hypervolume of each qNEHVI search as a percentage of its value after 2,304 evaluations (higher is better). Each endpoint-targeted search evaluated 3,200 policies, while the qNEHVI search evaluated 2,304 policies. Marginal improvement is small for later batches.}
    \label{fig:policyeval}
\end{figure}

\Cref{fig:policyeval} provides optimization convergence heuristics, and \Cref{tab:borough-sla-outcomes} reports the realized median inspection delays and inspection fractions under historical inspections and the selected Borough-budget policies.

\begin{table}[p]
\centering
\small
\setlength{\tabcolsep}{3pt}
\renewcommand{\arraystretch}{0.93}
\begin{tabular}{@{}lrrrrrrrr@{}}
\toprule
\multicolumn{1}{c}{\multirow{2}{*}{Category}} & \multicolumn{2}{c}{Historical} & \multicolumn{2}{c}{Most Efficient} & \multicolumn{2}{c}{Most Equitable} & \multicolumn{2}{c}{Balanced}\\ \cmidrule(l){2-9}
\multicolumn{1}{c}{} &
  \multicolumn{1}{c}{\begin{tabular}[c]{@{}c@{}}Delay\\ (days)\end{tabular}} &
  \multicolumn{1}{c}{\begin{tabular}[c]{@{}c@{}}Inspection\\ fraction\end{tabular}} &
  \multicolumn{1}{c}{\begin{tabular}[c]{@{}c@{}}Delay\\ (days)\end{tabular}} &
  \multicolumn{1}{c}{\begin{tabular}[c]{@{}c@{}}Inspection\\ fraction\end{tabular}} &
  \multicolumn{1}{c}{\begin{tabular}[c]{@{}c@{}}Delay\\ (days)\end{tabular}} &
  \multicolumn{1}{c}{\begin{tabular}[c]{@{}c@{}}Inspection\\ fraction\end{tabular}} &
  \multicolumn{1}{c}{\begin{tabular}[c]{@{}c@{}}Delay\\ (days)\end{tabular}} &
  \multicolumn{1}{c}{\begin{tabular}[c]{@{}c@{}}Inspection\\ fraction\end{tabular}}\\ \midrule
\multicolumn{9}{@{}l}{\textbf{Bronx}} \\
Hazard & 2 & 0.82 & 0.00 & 0.95 & 1.60 & 0.93 & 0.84 & 0.95 \\
Illegal Tree Damage & 14 & 0.81 & 0.00 & 0.99 & 0.00 & 0.97 & 0.00 & 1.00 \\
Other & 13 & 0.93 & 0.00 & 1.00 & 0.00 & 1.00 & 0.00 & 1.00 \\
Prune & 17 & 0.85 & 12.62 & 0.11 & 55.70 & 0.09 & 18.82 & 0.10 \\
Remove Tree & 7 & 0.90 & 3.16 & 0.82 & 8.20 & 0.72 & 3.64 & 0.81 \\
Root/Sewer/Sidewalk & 22 & 0.56 & 12.38 & 0.14 & 1.20 & 0.93 & 11.80 & 0.12 \\
\addlinespace[1.5pt]
\multicolumn{9}{@{}l}{\textbf{Brooklyn}} \\
Hazard & 4 & 0.86 & 0.20 & 0.99 & 1.00 & 0.94 & 0.76 & 0.99 \\
Illegal Tree Damage & 10 & 0.91 & 0.00 & 1.00 & 0.00 & 1.00 & 0.00 & 1.00 \\
Other & 13 & 0.72 & 0.00 & 1.00 & 0.00 & 1.00 & 0.00 & 1.00 \\
Prune & 8 & 0.17 & 17.92 & 0.18 & 26.20 & 0.14 & 20.88 & 0.16 \\
Remove Tree & 18 & 0.94 & 4.72 & 0.94 & 9.08 & 0.71 & 6.80 & 0.84 \\
Root/Sewer/Sidewalk & 28 & 0.66 & 18.20 & 0.29 & 0.00 & 1.00 & 17.76 & 0.26 \\
\addlinespace[1.5pt]
\multicolumn{9}{@{}l}{\textbf{Manhattan}} \\
Hazard & 1 & 0.73 & 1.00 & 0.97 & 1.08 & 0.95 & 1.00 & 0.97 \\
Illegal Tree Damage & 4 & 0.50 & 0.04 & 1.00 & 1.00 & 0.99 & 0.00 & 1.00 \\
Other & 3 & 0.92 & 0.16 & 1.00 & 0.00 & 0.99 & 0.00 & 1.00 \\
Prune & 8 & 0.86 & 10.14 & 0.11 & 15.40 & 0.14 & 17.08 & 0.28 \\
Remove Tree & 3 & 0.63 & 9.44 & 0.55 & 9.20 & 0.72 & 2.72 & 0.79 \\
Root/Sewer/Sidewalk & 22 & 0.15 & 0.00 & 0.99 & 0.00 & 1.00 & 0.00 & 0.99 \\
\addlinespace[1.5pt]
\multicolumn{9}{@{}l}{\textbf{Queens}} \\
Hazard & 1.50 & 0.84 & 2.44 & 0.99 & 0.00 & 0.92 & 1.40 & 0.99 \\
Illegal Tree Damage & 26 & 0.73 & 0.00 & 1.00 & 0.00 & 1.00 & 0.00 & 1.00 \\
Other & 7 & 0.81 & 0.00 & 0.99 & 0.00 & 1.00 & 0.00 & 0.99 \\
Prune & 8 & 0.24 & 19.20 & 0.05 & 16.60 & 0.08 & 16.40 & 0.07 \\
Remove Tree & 7 & 0.84 & 2.00 & 0.77 & 2.12 & 0.64 & 2.00 & 0.76 \\
Root/Sewer/Sidewalk & 22 & 0.64 & 23.42 & 0.07 & 5.72 & 0.11 & 20.40 & 0.11 \\
\addlinespace[1.5pt]
\multicolumn{9}{@{}l}{\textbf{Staten Island}} \\
Hazard & 1 & 0.70 & 1.00 & 1.00 & 0.48 & 0.95 & 1.00 & 0.99 \\
Illegal Tree Damage & 6 & 0.51 & 6.84 & 0.84 & 0.00 & 1.00 & 0.00 & 1.00 \\
Other & 5 & 0.78 & 0.00 & 0.99 & 0.00 & 0.99 & 0.00 & 0.99 \\
Prune & 3 & 0.38 & 6.42 & 0.14 & 25.24 & 0.11 & 12.64 & 0.12 \\
Remove Tree & 3 & 0.81 & 2.24 & 0.93 & 7.96 & 0.70 & 5.44 & 0.82 \\
Root/Sewer/Sidewalk & 27 & 0.69 & 5.48 & 0.21 & 25.08 & 0.14 & 17.90 & 0.18 \\
\bottomrule
\end{tabular}
\caption{Realized median inspection delays and inspection fractions by Borough and category for historical inspections and three selected Borough-budget policies. Simulated values are means across 25 simulations.}
\label{tab:borough-sla-outcomes}
\end{table}

\subsubsection{Sensitivity to category priority weights.}
\label{app:robustness}
\label{app:priority-weight-sensitivity}

Here, we hold the main policies fixed and assess their performance under alternative category priority weights. Our main analysis uses the assigned category priority weights reported in
App. \Cref{tab:avgrisk}. One concern with any choice of weights, even those with expert feedback, is that they do not correspond to the objective function of interest -- and the performance of the chosen policies and the comparison to historical outcomes are not robust. We thus conduct a sensitivity analysis. We hold the most efficient, most
equitable, and Balanced Borough budget policies fixed (from \Cref{tab:robustnesstab}) and recalculate their efficiency and equity losses under alternative priority weights. For every specification, we also recalculate historical loss using the same weights and report the ratio of policy loss to historical loss. This does not require resimulation, only rescoring simulation outcomes under a new loss; thus, we can do so efficiently, without reoptimizing or resimulating -- however, this approach means that the ``most efficient/equitable'' policies are optimized for a misspecified objective (the primary assigned weights).

We consider two sets of weights. We first consider interpretable weights: equal weights; the mean recorded risk in 2019--2022 for each
category--Borough pair; and several order-preserving transformations of the
assigned category weights, which raise each assigned category weight to the
power $\eta$ -- lower $\eta$ corresponds to more equal weights than the main assigned specification, and higher to more extreme weights (prioritizing higher risk categories even more). We then consider random weights, sampling 10,000 sets of weights by drawing one independent Uniform$(0,1)$ value for each of the four distinct assigned-priority levels, sorting these values, and preserving the assigned ordering and ties across categories. (In all cases, we normalize the mean weights, but this does not affect the ratio evaluations.)

\begin{table}[tb]
\centering
\small
\setlength{\tabcolsep}{3pt}
\resizebox{\textwidth}{!}{\begin{tabular}{@{}lrrrrrrr@{}}
\toprule
& \multicolumn{2}{c}{Most Efficient} & \multicolumn{2}{c}{Most Equitable} & \multicolumn{2}{c}{Balanced} & \\
\cmidrule(lr){2-3}\cmidrule(lr){4-5}\cmidrule(lr){6-7}
Priority-weight specification & Efficiency & Equity & Efficiency & Equity & Efficiency & Equity & Price of equity \\
\midrule
Assigned & 0.800 & 0.825 & 0.902 & 0.426 & 0.808 & 0.447 & 12.8\% \\
Equal & 1.021 & 0.858 & 1.025 & 0.552 & 1.026 & 0.589 & 0.4\% \\
Average recorded risk & 0.964 & 0.813 & 1.015 & 0.512 & 0.971 & 0.565 & 5.3\% \\
Power $\eta=0.25$ & 0.968 & 0.849 & 0.997 & 0.520 & 0.974 & 0.553 & 2.9\% \\
Power $\eta=0.5$ & 0.913 & 0.841 & 0.966 & 0.488 & 0.920 & 0.517 & 5.8\% \\
Power $\eta=0.75$ & 0.857 & 0.833 & 0.935 & 0.457 & 0.864 & 0.481 & 9.1\% \\
Power $\eta=1.25$ & 0.744 & 0.817 & 0.870 & 0.397 & 0.753 & 0.414 & 16.9\% \\
Power $\eta=1.5$ & 0.690 & 0.810 & 0.838 & 0.370 & 0.700 & 0.384 & 21.5\% \\
Power $\eta=2$ & 0.590 & 0.796 & 0.777 & 0.322 & 0.601 & 0.331 & 31.6\% \\
\midrule
Sampled 1st--99th percentile & [0.290, 0.967] & [0.638, 0.892] & [0.549, 0.994] & [0.148, 0.527] & [0.298, 0.972] & [0.130, 0.563] & --- \\
Sampled componentwise maximum & 1.013 & 0.930 & 1.021 & 0.548 & 1.018 & 0.584 & --- \\
\bottomrule
\end{tabular}}
\caption{Priority-weight sensitivity analyses. Efficiency and equity entries are policy loss divided by historical loss under the same weights. The price of equity is the proportional increase in efficiency loss from the most efficient to the most equitable fixed policy. In short, we find that equity gains from our optimized policies are more robust across weights, while efficiency gains are small when the category weights are more equal across categories but the policies generally improve on historical inspections.}
\label{tab:priority-weight-sensitivity}
\end{table}

Results are shown in \Cref{tab:priority-weight-sensitivity}. Several patterns emerge. The optimized policies outperform historical outcomes on both metrics in almost every set of weights, including over 99\% of random weights. The exceptions largely occur when category weights are near-equal across types (also corresponding more closely to average recorded risk), in which case the efficiency losses of our policies are approximately equal to the loss under historical outcomes (the ratios are about $1.02$). Furthermore, the gain in equity is more consistent and generally larger than the efficiency gain -- our most efficient endpoint improves on historical performance by about 20\%, while the most equitable policy gains about 50\%. Consistent with our other results, these analyses suggest that historical outcomes are more consistent with treating all incidents similarly for inspection allocation, regardless of category, and prioritizing efficiency over equity -- this is perhaps interesting given the practitioner feedback against more equal category weights. Relatedly, the price of equity rises as the category weights become more extreme, suggesting that tradeoffs between Boroughs are partially driven by differences in category distributions and how those correspond to budget allocations.

\subsubsection{Sensitivity to objective specifications and inspection capacity.}
\label{app:altobj}
\label{app:times2budget}

\begin{figure}[!htbp]
    \centering
    \includegraphics[width=0.95\textwidth]{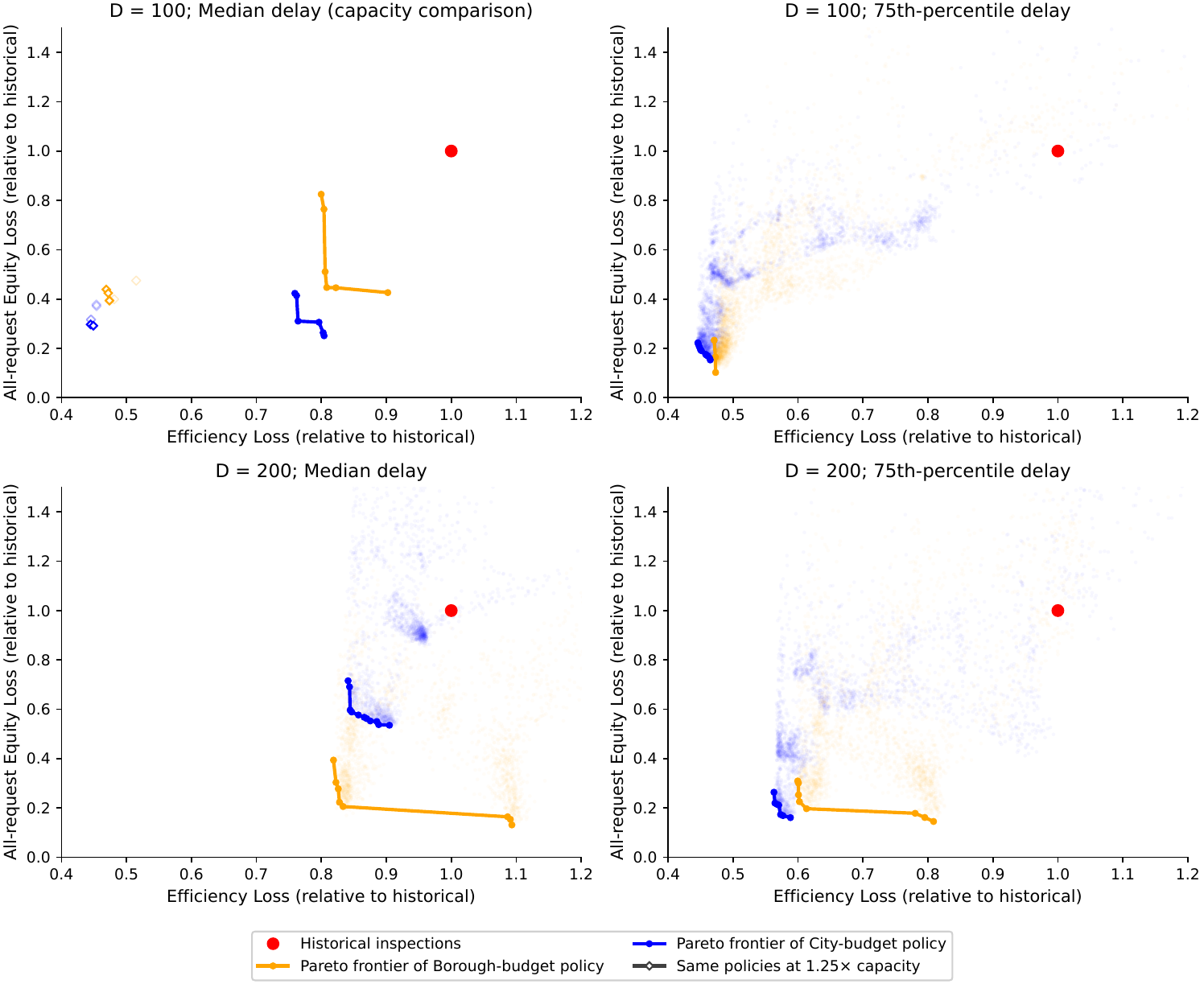}
    \caption{Robustness of the equity--efficiency trade-off conceptual findings under different specifications. The first subplot evaluates the policies from the main analysis at historical capacity (circles) and 1.25 times historical daily capacity (diamonds). As expected, higher capacity substantially reduces the tradeoff {in absolute terms}. The remaining three figures show separately optimized
    policy frontiers under different choices of percentile of delay ($1-\alpha$) or the cost of dropping incidents ($D$). Faint dots are individual Bayesian-optimization
    evaluations, each based on one simulation. Pareto-optimal policies are evaluated across 25 simulations.
     Orange and blue indicate Borough- and City-budget policies, respectively, and red indicates historical inspections in 2019. Each alternative-specification panel uses its own historical losses as the denominators, while the capacity panel uses the historical losses from the main analysis, so history appears at $(1,1)$ in every panel.}
    \label{fig:pfvarsla}
    \label{fig:paretotimes2}
\end{figure}

This subsection examines sensitivity to alternative delay statistics,
noninspection penalties, and inspection capacity. Our main analysis defines performance using median inspection delays, assigns a cost of $D=100$ to noninspection, and uses historical daily inspection capacity. We examine three alternative objective specifications: 75th-percentile delay with $D=100$, and median or 75th-percentile delay with $D=200$. For each specification, we reoptimize the Borough- and City-budget policies and evaluate the resulting policies across 25 simulations. \Cref{fig:pfvarsla} (top right and bottom row) shows that both policy classes continue to contain policies that reduce both efficiency and equity loss relative to historical inspections. The relative performance of the two policy classes varies across specifications. Thus, the possibility of improving both objectives is robust, while the magnitude of the trade-off and the gain from centralization depend on the objective specification.

We also examine the role of inspection capacity. As suggested by
\Cref{sec:equity_efficiency_tradeoff}, the magnitude of the equity--efficiency
trade-off {in absolute terms} depends on capacity slack. In the top left subplot of
\Cref{fig:paretotimes2}, we simulate the policies from the main analysis using
historical daily inspection capacity and 1.25 times that capacity. The
trade-off among these policies is substantially smaller {in absolute terms},
consistent with the capacity-slack result {for the absolute price of equity} in the theoretical model. (Note that because we evaluate the same policies at both capacity levels, this comparison isolates how greater capacity changes their performance; it does not characterize the frontier that would result from reoptimizing policies at the higher capacity).

\FloatBarrier

\section{Omitted proofs}
\begingroup
\label{app:fixed-load-theory-proofs}

Throughout, $\mathcal S$ and $\mathcal B$ are finite,
$\lambda_{k,b}>0$, $0<s_{k,b}\leq\lambda_{k,b}$, $r_{k,b}>0$,
$D_{k,b}\geq0$, and $E(s)>0$. Let $a_{k,b}=-\log\alpha_{k,b}>0$.

\subsection{Proof of \Cref{prop:opt-reformulation}}
\reformulation*
We prove \Cref{prop:opt-reformulation} using general $a_{k,b}$, though the proofs below will use the main-text assumption $a_{k,b} = a$. 
The original fixed-load program is convex; constraints other than Eq.~\eqref{opt-sla} are linear, and $L(z)$ is assumed convex. Constraint~\eqref{opt-sla} is convex, as follows. Rewrite it as $0\geq s_{k,b}/\phi_{k,b}+a_{k,b}/(\phi_{k,b}z_{k,b}) - C_b$. The Hessian of the right-hand side with respect to $(\phi_{k,b}, z_{k,b}, C_b)$ is positive semi-definite. Next, we show equivalence of the constraints.

Summing over the constraints in \eqref{opt-sla} and applying Eqs.~\eqref{opt-gps} and \eqref{opt-budget}, every feasible vector $z$ satisfies
\begin{equation}
 \sum_{k,b}\frac{a_{k,b}}{z_{k,b}}
 \leq \sum_b C_b\sum_k\phi_{k,b}-\sum_{k,b}s_{k,b}
 \leq C-\sum_{k,b}s_{k,b}\triangleq E(s).
\end{equation}
And so Eqs.~\eqref{opt-sla}--\eqref{opt-budget} imply the constraint in Eq.~\eqref{opt-reformulated}.
Conversely, if $z>0$ satisfies the constraint in \eqref{opt-reformulated}, define
\begin{equation*}
 C_b=\sum_k\left(s_{k,b}+\frac{a_{k,b}}{z_{k,b}}\right),
 \qquad
 \phi_{k,b}=\frac{s_{k,b}+a_{k,b}/z_{k,b}}{C_b}.
\end{equation*}
Then $\sum_k\phi_{k,b}=1$, $\sum_bC_b\leq C$, and every SLA constraint
binds. When the reciprocal-capacity constraint binds, the construction is unique
given $z$.  Indeed, any supporting policy has
$C_b\phi_{k,b}\geq s_{k,b}+a_{k,b}/z_{k,b}$, and hence
\begin{equation*}
 C=\sum_{k,b}\left(s_{k,b}+\frac{a_{k,b}}{z_{k,b}}\right)
 \leq\sum_b C_b\sum_k\phi_{k,b}
 \leq\sum_bC_b\leq C.
\end{equation*}

\subsection{Proof of \Cref{prop:extremeeff}}
\propextremeefficiency*
The most efficient solution minimizes
\begin{align*}
G(s,z)
&=\sum_{k,b}s_{k,b}r_{k,b}z_{k,b}
  +\sum_{k,b}(\lambda_{k,b}-s_{k,b})r_{k,b}D_{k,b}
\end{align*}
given the constraint in Eq.~\eqref{opt-reformulated}, and recall that
$A_{\mathrm{eff}}(s)=\sum_{k,b}\sqrt{a s_{k,b}r_{k,b}}$. The second term of $G$ (corresponding to the noninspection component) is fixed given $s$ and can be ignored.  Cauchy--Schwarz gives
\begin{equation}
 \left(\sum_{k,b}s_{k,b}r_{k,b}z_{k,b}\right)
 \left(\sum_{k,b}\frac{a}{z_{k,b}}\right)
 \geq A_{\mathrm{eff}}(s)^2,
\end{equation}
and equality holds exactly when $z_{k,b}\sqrt{s_{k,b}r_{k,b}/a}$ is constant, or, equivalently, $z_{k,b} = c \sqrt{\frac{a}{s_{k,b}r_{k,b}}}$ for some $c$. Note that the constraint is binding, and so $\sum_{k,b}\frac{a}{z_{k,b}} = E(s)$; solving for $c$ yields the solution.

\subsection{Proof of \Cref{prop:extremefair}}
\propextremeequity*
The most equitable solution minimizes
\begin{equation*}
 F(s,z)=\sum_k\left[\max_b\operatorname{Cost}_{k,b}(s,z)
 -\min_b\operatorname{Cost}_{k,b}(s,z)\right] \geq0
\end{equation*}
given the constraint in Eq.~\eqref{opt-reformulated}.  Write $\operatorname{Cost}_{k,b}(s,z)=c_{k,b}z_{k,b}+d_{k,b}$, where
\begin{equation*}
 c_{k,b}=\frac{s_{k,b}r_{k,b}}{\lambda_{k,b}}>0,
 \qquad
 d_{k,b}=\left(1-\frac{s_{k,b}}{\lambda_{k,b}}\right)r_{k,b}D_{k,b}.
\end{equation*}  For any
$M_k>\max_b d_{k,b}$, setting $z_{k,b}=\frac{M_k-d_{k,b}}{c_{k,b}}$
gives $\operatorname{Cost}_{k,b}(s,z)=M_k$.  As every $M_k$ increases, so does $z_{k,b}$. Thus,
$\sum_{k,b}a/z_{k,b}$ converges to zero, so these thresholds can be chosen to
satisfy Eq.~\eqref{opt-reformulated}.  Thus the minimum equity loss is zero,
which proves the stated categorywise equality. Note that we can then show that the efficiency-best solution among solutions equalizing costs exists, due to continuity of $G$ and compactness in $z$ (which is lower bounded due to the feasibility constraint; then, a single solution to the equity minimizing optimization produces a sublevel set for $G$, bounding $z$ from above.) Finally, note that the constraint binds at the efficiency-best equitable solution (otherwise, we can perturb $z$ to improve the efficiency loss while keeping the equity loss at $0$).

\subsection{Proof of \Cref{prop:costofequity}}

\fixedloadalignment*
By the proofs of \Cref{prop:extremeeff} and \Cref{prop:extremefair}, both $z^{\mathrm{eff}}$ and $z^{\mathrm{eq}}$ bind the constraint.
For any binding $z$, recall that
\begin{equation*}
 q_{k,b}(z)=\frac{a}{E(s)z_{k,b}}
\end{equation*}
is the fraction of total capacity slack $E(s)$ assigned to cell $(k,b)$; hence
$q(z)$ is a probability vector.  For probability vectors $P$ and $Q$, recall
that the Pearson chi-square divergence is
$\chi^2(P\|Q)=\sum_{k,b}(P_{k,b}-Q_{k,b})^2/Q_{k,b}$.
By \Cref{prop:extremeeff}, we have
$q^{\mathrm{eff}}_{k,b}
 =\frac{\sqrt{a s_{k,b}r_{k,b}}}{A_{\mathrm{eff}}(s)}.$ 
Recall also that $G(s,z)
=\sum_{k,b}s_{k,b}r_{k,b}z_{k,b} +\sum_{k,b}(\lambda_{k,b}-s_{k,b})r_{k,b}D_{k,b}.$ 
The second sum is the fixed noninspection component, which does not depend on
$z$.

Now consider the inspection component. For any
binding $z$, the definitions of $q(z)$ and $q^{\mathrm{eff}}$ give
\begin{equation*}
 z_{k,b}=\frac{a}{E(s)q_{k,b}(z)},
 \qquad
 s_{k,b}r_{k,b}
 =\frac{A_{\mathrm{eff}}(s)^2}{a}(q^{\mathrm{eff}}_{k,b})^2.
\end{equation*}
Multiplying these identities cell by cell and summing, the inspected-delay
component satisfies
\begin{align*}
\sum_{k,b}s_{k,b}r_{k,b}z_{k,b}
&=\frac{A_{\mathrm{eff}}(s)^2}{E(s)}
 \sum_{k,b}\frac{(q^{\mathrm{eff}}_{k,b})^2}{q_{k,b}(z)}\\
&=\frac{A_{\mathrm{eff}}(s)^2}{E(s)}
 \left[1+\chi^2\!\left(q^{\mathrm{eff}}\middle\|q(z)\right)\right].
\end{align*}
The last equality uses
$\sum_iP_i^2/Q_i=1+\chi^2(P\|Q)$.  Evaluating the identity at the two
endpoints and subtracting gives
\begin{equation*}
 G(s,z^{\mathrm{eq}})-G(s,z^{\mathrm{eff}})
 =\frac{A_{\mathrm{eff}}(s)^2}{E(s)}
 \chi^2\!\left(q^{\mathrm{eff}}\middle\|q^{\mathrm{eq}}\right).
\end{equation*}
Dividing by $G(s,z^{\mathrm{eff}})>0$ proves Eq.~\eqref{eq:price-of-equity-chi-square}.
The price is zero exactly when the divergence is zero, or equivalently when
$q^{\mathrm{eq}}=q^{\mathrm{eff}}$ and hence when the endpoint
thresholds coincide.  Then, \Cref{prop:extremefair} implies that the
efficient cell costs are constant across Boroughs within each category.
Conversely, if the efficient cell costs are already constant, then
$z^{\mathrm{eff}}$ is equitable and globally efficiency minimizing, so it is
the efficiency-best equitable solution and coincides with $z^{\mathrm{eq}}$.

For the upper bound, let
$\overline M_k=\max_{b'}\operatorname{Cost}_{k,b'}(s,z^{\mathrm{eff}})$ and
raise each $z^{\mathrm{eff}}_{k,b}$ to the unique $\widetilde z_{k,b}$ such that
$\operatorname{Cost}_{k,b}(s,\widetilde z)=\overline M_k$.  Then
$\widetilde z\geq z^{\mathrm{eff}}$, so it remains feasible, and
$F(s,\widetilde z)=0$.  The efficiency tie-break therefore gives
\begin{align*}
 \operatorname{PoE}_C(s)
 &\leq \frac{G(s,\widetilde z)-G(s,z^{\mathrm{eff}})}
 {G(s,z^{\mathrm{eff}})}\\
 &=\frac{1}{G(s,z^{\mathrm{eff}})}\sum_{k,b}\lambda_{k,b}\left[
 \overline M_k-\operatorname{Cost}_{k,b}(s,z^{\mathrm{eff}})\right].
\end{align*}

Finally, under the alignment assumption, write
$d_k=r_{k,b}D_{k,b}(1-s_{k,b}/\lambda_{k,b})$, which is independent of $b$,
and let
\begin{equation*}
 t=\frac{C-\sum_{k,b}s_{k,b}}{C'-\sum_{k,b}s_{k,b}}.
\end{equation*}
The map $z\mapsto tz$ is a bijection between the feasible sets at $C$ and
$C'$, since $\sum_{k,b}a/(tz_{k,b})=t^{-1}\sum_{k,b}a/z_{k,b}$.  Moreover,
\begin{equation*}
 \operatorname{Cost}_{k,b}(s,tz)
 =t\operatorname{Cost}_{k,b}(s,z)+(1-t)d_k,
\end{equation*}
and so both the efficiency and equity solutions translate.  Consequently,
\begin{align*}
&G(s,z^{\mathrm{eq}}(C'))-G(s,z^{\mathrm{eff}}(C'))\\
&\qquad=t\left[G(s,z^{\mathrm{eq}}(C))-G(s,z^{\mathrm{eff}}(C))\right].
\end{align*}
Dividing by $G(s,z^{\mathrm{eff}}(C'))$ proves
Eq.~\eqref{eq:poe-capacity-scaling}.  Finally, by \Cref{prop:extremeeff},
\begin{equation*}
G(s,z^{\mathrm{eff}}(C))
=\frac{A_{\mathrm{eff}}(s)^2}{C-\sum_{k,b}s_{k,b}}
+\sum_{k,b}(\lambda_{k,b}-s_{k,b})r_{k,b}D_{k,b}.
\end{equation*}
Substitution shows that the multiplicative factor in
Eq.~\eqref{eq:poe-capacity-scaling} is less than one when the second sum is
positive and equals one when it is zero.

\subsection{Proof of \Cref{prop:strict_improvements_from_centralization}}
\propcentralized*
The city-efficient solution can be derived in the same way as the Borough-efficient solution in \Cref{prop:extremeeff}, yielding
\begin{equation*}
 z^{\mathrm{city}}_k
 =\frac{A^{\mathrm{city}}(s)}{E(s)}
   \sqrt{\frac{a}{\sum_b s_{k,b}r_{k,b}}}.
\end{equation*}
Using \Cref{prop:extremeeff},
\begin{equation*}
 \frac{z^{\mathrm{city}}_k}{z^{\mathrm{eff}}_{k,b}}
 =\frac{A^{\mathrm{city}}(s)}{A_{\mathrm{eff}}(s)}
   \sqrt{\frac{s_{k,b}r_{k,b}}
   {\sum_{b'}s_{k,b'}r_{k,b'}}}<1.
\end{equation*}
For each category,
$\sqrt{\sum_bx_b}\leq\sum_b\sqrt{x_b}$, so
$A^{\mathrm{city}}(s)\leq A_{\mathrm{eff}}(s)$; the square-root factor is
strictly less than one because all cells are positive and $|\mathcal B|\geq2$.

The optimized variable response-cost terms are
$A_{\mathrm{eff}}(s)^2/E(s)$ and $A^{\mathrm{city}}(s)^2/E(s)$, while the
fixed noninspection terms agree.  Subtracting yields the numerator in Eq.~\eqref{eq:aligned-pooling-gain}, and then dividing by
$G(s,z^{\mathrm{eff}})$ finishes.  Since the relative
centralization gain and price of equity have the same positive denominator,
that denominator cancels in their comparison.  Using the capacity-share
vectors $q$ from the proof of \Cref{prop:costofequity}, we have
\begin{align*}
&\textnormal{centralization gain}
 \geq\textnormal{price of equity}\\
&\Longleftrightarrow\quad
\frac{A_{\mathrm{eff}}(s)^2-A^{\mathrm{city}}(s)^2}{E(s)}
\geq\frac{A_{\mathrm{eff}}(s)^2}{E(s)}
\chi^2\!\left(q^{\mathrm{eff}}\middle\|q^{\mathrm{eq}}\right)\\
&\Longleftrightarrow\quad
1-\frac{A^{\mathrm{city}}(s)^2}{A_{\mathrm{eff}}(s)^2}
\geq\chi^2\!\left(q^{\mathrm{eff}}\middle\|q^{\mathrm{eq}}\right).
\end{align*}

\endgroup

\end{document}